\documentclass[twocolumn,floats,floatfix,showpacs,amsmath,amssymb,prd,superscriptaddress,nofootinbib]{revtex4-1}

\usepackage{color}

\usepackage{hyperref}
\usepackage{graphicx}
\usepackage{graphics}
\usepackage{epsfig}
\usepackage{bm}
\usepackage{longtable}                 
\usepackage{verbatim}
\usepackage{yfonts}
\usepackage{psfrag}
\usepackage{url}
\usepackage{subfigure}
\usepackage{color}
\input{epsf} 
\usepackage{thumbpdf}
\usepackage{pdfsync}
\usepackage{graphicx}
\def\be{\begin{equation}}
\def\en{\end{equation}}
\def\bea{\begin{eqnarray}}
\def\ena{\end{eqnarray}}

\def\msun{\,{\rm M_\odot}}

\begin{document}

\title{Resolving multiple supermassive black hole binaries with pulsar timing arrays}

\author{Stanislav Babak} 
\email{Stanislav.Babak@aei.mpg.de}
\affiliation{Albert Einstein Institute, Am Muhlenberg 1 D-14476 Golm, Germany}
\author{Alberto Sesana} 
\email{Alberto.Sesana@aei.mpg.de}
\affiliation{Albert Einstein Institute, Am Muhlenberg 1 D-14476 Golm, Germany}

\date{\today}

\begin{abstract}
We study the capability of a pulsar timing array (PTA) to individually resolve and localize in 
the sky monochromatic gravitational wave (GW) sources. Given a cosmological population
of inspiralling massive black hole binaries, their observable signal in the PTA domain
is expected to be a superposition of several nearly-monochromatic GWs of different strength.
In each frequency bin, the signal is neither a stochastic background nor perfectly resolvable in its
individual components. In this context, it is crucial to explore how the information encoded in the
spatial distribution of the array of pulsars might help recovering the origin of the GW signal, by resolving
individually and locating in the sky the strongest sources.  In this paper we develop a maximum-likelihood
based method finalized to this purpose.  We test the algorithm against
noiseless data showing that up to $P/3$ sources can be resolved and localized in the sky by an array
of $P$ pulsars. We validate the code by performing blind searches on both noiseless and noisy data 
containing an unknown number of signals with different strengths. Even without employing any proper
search algorithm, our analysis procedure performs well, recovering and correctly locating in the sky 
almost all the injected sources. 
\end{abstract}

 

\maketitle

\section{Introduction}
Within this decade the detection of gravitational waves (GWs) may be a reality,
opening a completely new window on the Universe.  Precision timing of millisecond 
pulsars  within a pulsar timing array (PTA) provides a unique opportunity to get 
the very first low-frequency GW detection. GWs affect the propagation of radio signals 
from the pulsar to the receiver on Earth, leaving a 
characteristic fingerprint in the time of arrival of the radio pulses
 \cite[e.g.,][]{sazhin78,hellings83}. This technique is particularly
sensitive in the $10^{-9}-10^{-7}$Hz window, where genuinely supermassive black hole
binaries (SMBHBs, $10^8-10^9\msun$) at low redshift ($z<2$) will dominate the signal 
(Sesana et al.  2009). The Parkes Pulsar Timing Array \cite[PPTA,~][]{manchester2008}, 
the European Pulsar Timing Array \cite[EPTA,~][]{janssen2008} and the North American 
Nanohertz Observatory for Gravitational Waves \cite[NANOGrav,~][]{jenet2009}, joining 
together in the International Pulsar Timing Array \cite[IPTA,~][]{hobbs2010}, are already collecting
data and improving their sensitivity in the frequency range of $\sim10^{-9}-10^{-6}$
Hz. In the coming years, the Chinese five hundred meter aperture spherical telescope
\cite[FAST,~][]{smits09} and the planned Square Kilometer Array
\cite[SKA,~][]{lazio2009} will provide a major leap in sensitivity.  

The signal expected from a cosmological population of inspiralling SMBHBs
consists of a superposition of quasi-monochromatic waves, similar to the white dwarf-white dwarf
foreground  \cite[e.g.,][]{Nelemans01} in the mHz window relevant to space based 
interferometry. Such signal can be divided into two distinct contributions: (i) 
a stochastic background generated by the incoherent superposition of radiation 
from the whole SMBHB population \cite{rr95,jaffe03,sesana08} and (ii) 
individually resolvable, deterministic signals produced by single sources that are 
sufficiently massive and/or close so that the gravitational signal stands above the 
root-mean-square (rms) level of the background \cite{sesana09}. 
Accordingly, signal recovery and data analysis investigations
have been developed in the limiting cases of either an isotropic stochastic background
described by a power-law spectrum \cite{jen05,jen06,anholm09, vanh09, vanh11}, or of 
a single GW source \cite{jen04,SesanaVecchio10,corbin2010,yardley10,lee11}.  

The actual expected signal, however, is far from being isotropically distributed in the sky, 
with just few sources dominating the power at each frequency \cite{sesana08,KS11}. The GW 
signal generated by an astrophysically motivated SMBHB population is neither isotropic and 
stochastic, nor dominated by a single source. It is therefore important to fully 
characterize such signal, and to assess how much of the information enclosed in it can be 
recovered. It is, in fact, not clear a priori how such inhomogeneous combination
of multiple sources emitting in the same frequency bin will appear in the time residuals: 
whether they will look almost isotropic over the sky, or they will superpose 
and appear as a single one, or as few bright spots in the sky.

This paper is a proof of concept of how the spatial information enclosed in a PTA can
be exploited to disentangle a superposition of signals coming from different directions in the 
sky. We utilize a maximum-likelihood based algorithm designed to identify the sky origin
and relative strengths of a collection of incoherently superposed monochromatic signals. We 
test the algorithm performances on different noiseless datasets, and then we apply it to
noisy datasets. In agreement with simple counting arguments and with
previous analytical results \cite{boyle10} we find that $P/3$ sources can be resolved and 
localized in the sky by an array of $P$ pulsars. 
If the number of sources is more than $P/3$ we might identify the presence 
of extra sources with poor localization precision, and we will not be able to estimate 
their parameters. When few sources are present,
the maximum likelihood estimator allows to infer the correct number of contributing 
sources, and the algorithm locates them in the sky with good accuracy. The same is true
when noise is added to the data. Also in a situation of source SNR$\approx 10$, the algorithm
is able to correctly disentangle signals and to locate the sources in the sky. 

This pilot study demonstrates the potential of PTAs of resolving and locating 
in the sky a sizable number of sources. We stress that we do not implement any proper 
search algorithm when looking for solutions that maximize the likelihood estimator, 
and we do not draw sources from expected astrophysical SMBHB populations. We plan to consider 
these aspects of the problem in the future, to test our method under more realistic assumptions. 

The paper is organized as follows. We spell out the objectives and the assumptions of our data 
analysis method in Section \ref{S:objectives}. The model used for the sources and a detailed
mathematical description of the detection strategy can be found in Section \ref{S:GWmodel}.
We discuss source identification and resolvability in Section \ref{resolvability}
and test our method against blind datasets in Section \ref{blind}.
We summarize and discuss future research directions in Section \ref{discussion}.

\section{Objectives and basic assumptions}
\label{S:objectives}

The main aim of this paper is to provide a proof of concept of the potential of PTAs to 
resolve individual contributions to a GW signal formed by the incoherent superposition 
of several monochromatic sources. We are primarily interested here in disentangling  the source 
contribution to the signal and in estimating their sky location. As we aim at a proof 
of concept, we make a set of simplifying assumptions: 

\begin{enumerate}

\item only GW sources contribute to the timing residuals, i.e., we test our method on noiseless data.
We will relax this assumption in the last Section, testing the effectiveness of our technique
on noisy synthetic data. When included, noise is modelled as white and Gaussian, with equal 
magnitude in all the pulsars forming the array;

\item we only consider few GW sources contributing to the data streams. We will consider 
more realistic source populations where few bright sources might dominate over a large distribution 
of weaker ones in future investigations;

\item we look at signals in a particular `frequency bin', i.e, we do not perform any frequency search. 
Throughout the paper we deal with \emph{monochromatic} signals at $f = 14$ nHz ($f=20$ nHz for 
the blind challenges). We assume that the frequency is known. We 
intend  to include frequency scanning in the analysis in forthcoming studies;

\item when modelling the GW signal, we use the leading quadrupole approximation for 
non-spinning binaries in quasi-circular orbit, as we will describe in the details in 
Section~\ref{S:GWmodel};

\item we consider the ``Earth'' term only in the pulsar response, excluding the ``pulsar'' term. 
GW sources aligned along the Earth-pulsar line of sight do not contribute 
to the timing residuals if the full response is considered (Earth$+$pulsar terms), however 
excluding the ``pulsar'' term makes the response infinite in this case. We 
therefore exclude GW sources within 5 degrees of any pulsar sky location.  
We will give more details on the ``pulsar'' and the ``Earth'' terms in Section~\ref{S:GWmodel};

\item the sky location of the pulsars in the array is drawn from a uniform distribution on the sphere;

\item we assume 10 years of uninterrupted equally-sampled (once every two weeks) observations of each pulsar 
in the array;

\item we do not implement any proper search algorithm in maximizing the likelihood. The sky maps and maximum 
likelihood estimates are computed by randomly (uniformly on the sphere) drawing points in the sky.
We will optimize the search algorithm in future investigations.

\end{enumerate}

Each residual series is monochromatic in our set up, and therefore contains two measurable 
quantities: the amplitude and the initial phase. On the other hand, in the approximation
of circular non evolving sources, the GW signal is described by six parameters: 
overall amplitude ${\cal A}$, inclination $\iota$, initial orbital phase $\Phi_0$, 
polarization angle $\psi$, sky location angles $\{\phi, \theta\}$. The simple counting of 
the measurable quantities (two per pulsar) and of the number of unknowns (six per GW source) suggests 
that we will need $P=3\times N${\footnote{Throughout the paper we will denote with $P$ the number
of pulsars in the array and with $N$ the number of sources contributing to the signal.}} 
pulsars to detect and to estimate parameters of $N$ GW sources. 
In general this counting is correct if one knows the exact number of signals in the data 
and wants to estimate parameters of all these signals. In a real situation 
the exact number of GW signals is unknown. In addition the strength of different GW sources 
could vary significantly, with a large fraction of weak sources forming an
anisotropic (in the sky) background for the brighter ones. 


\section{GW source model and detection strategy}
\label{S:GWmodel}

The analysis is built around the maximum likelihood estimator of the source parameters. We notice that the 
likelihood could be maximized {\it analytically} over some (what are called extrinsic) parameters 
characterizing the GW source, in a way similar to the $F$-statistic \cite{JKS}. We therefore need to write 
the signal in a mathematical form suitable to such analytical maximization.

\subsection{Gravitational wave model}
We consider monochromatic GW sources assuming that the orbital frequency does not 
change appreciably over the observation period \cite[see, e.g.,][]{SesanaVecchio10}
which we took to be 10 years. The signal is characterized by the six parameters 
introduced in the previous Section $\{{\cal A}, \iota, \Phi_0, \psi, \phi, \theta\}$, 
plus the GW frequency $f$, which we assume to be known here.
Those parameters describe the GW signal from a widely 
separated SMBHB in quasi-circular orbit. We neglect all modulations due to spin-orbital coupling
\cite{vecchio04}, so effectively we consider two non-spinning SMBHs. We refer the reader for more detailed discussion on the 
waveforms and on the computing the timing residuals to \cite{SesanaVecchio10}. The waveform in the radiative frame reads 
\bea
h_{ij} &=& \epsilon_{ij}^{+} h_{+} + \epsilon_{ij}^{\times} h_{\times},\\
\epsilon_{ij}^{+} &=& p_i p_j - q_i q_j,\;\;\; \epsilon_{ij}^{\times} = p_i q_j + p_j q_i ,
\ena
where $\hat{p}, \hat{q}$ is the polarization basis. The polarization basis associated to the Earth 
(or to the Solar system barycenter) is 
\bea
\hat{u} &=& \{\cos{\theta} \cos{\phi}, \cos{\theta} \sin{\phi}, -\sin{\theta} \},\\
\hat{v} &=& \{\sin{\phi}, -\cos{\phi}, 0 \},\\
\hat{p} &=& \hat{u}\cos{\psi}  - \hat{v}\sin{\psi}, \;\;\; \hat{q} = \hat{v}\cos{\psi}  + \hat{u}\sin{\psi}.
\ena
The rotation in the last line defines the polarization angle $\psi$.  The components of the GW signal are 
\bea
\label{eq:strain}
h_{+} &=& \mathcal{A} (1 + \cos^2{\iota}) \cos(\Phi(t) + \Phi_0)\nonumber\\
h_{\times} &=& - 2 \mathcal{A} \cos{\iota} \sin(\Phi(t) + \Phi_0)
\ena
where we have written explicitly the initial GW phase, and 
\bea 
\mathcal{A} = 2 \frac{\mathcal{M}_c^{5/3}}{D_L} (\pi f)^{2/3}.
\ena 
In the last equation $\mathcal{M}_c=M_1^{3/5}M_2^{3/5}/(M_1+M_2)^{1/5}$ is the chirp mass 
of the source, and $D_L$ is its luminosity distance. The residuals in the arrival time 
of the radio pulses due to the propagation of the electromagnetic waves in the field of the GW 
can be written as 
\bea
\label{eq:residual}
r(t) = \int_0^{t} \frac{\delta \nu}{\nu}(t') dt',\\
\frac{\delta \nu}{\nu} = \frac{1}{2} \frac{\hat{n}^i \hat{n}^j}{1 + \hat{n}\cdot\hat{k}} \Delta h_{ij},
\ena
where $\hat{n}$ denotes the position of the pulsar on the sky, and $\hat{k}$ is the direction of
the GW propagation. The last term depends on the strain of the GW at the location of the pulsar 
$h_{ij}(t_p)$ and at the Earth $h_{ij}(t)$:
\be
\Delta h_{ij} = h_{ij}(t_p) - h_{ij}(t).
\en
Since the pulsars are not correlated  ($t_p$, the emission time of the pulse detected at the time $t$
on the Earth, is different for all pulsars) the ``pulsar'' terms do not add up coherently 
and for this work we will neglect them and concentrate on the ``Earth'' terms only. The pulsar terms can
be considered as part of the noise, which for the time being we do not include in our analysis.  
Substituting equation (\ref{eq:strain}) into equation (\ref{eq:residual}), the 
Earth part of the residual is written as
\bea
r_{\alpha}^E(t) = \frac{\mathcal{A}}{2\pi f} \left\{   
(1 + \cos^2{\iota}) F_{\alpha}^{+} \left[ \sin(\Phi(t) + \Phi_0)  - \sin{\Phi_0}\right] + \right. \nonumber \\
\left. 2 \cos{\iota} F_{\alpha}^{\times} \left[ \cos(\Phi(t) + \Phi_0)  - \cos{\Phi_0}\right] 
\right\},
\ena
where the index $\alpha$ refers to a particular pulsar in the array, and the antenna-response functions are 
\bea
F_{\alpha}^{+} & = & \frac{1}{2} \frac{(\hat{n}^{\alpha}\cdot\hat{p})^2 -  (\hat{n}^{\alpha}\cdot\hat{q})^2}{1 + \hat{n}^{\alpha}\cdot\hat{k}}\\
F_{\alpha}^{\times} & = & \frac{(\hat{n}^{\alpha}\cdot\hat{p}) (\hat{n}^{\alpha}\cdot\hat{q})}{1 + \hat{n}^{\alpha}\cdot\hat{k}}.
\ena
It is convenient to rewrite the above expressions isolating the terms containing the polarization angle:
\bea
F_{\alpha}^{+} & = & F^{\alpha}_c \cos( 2\psi) + F^{\alpha}_s \sin(2\psi) \\
F_{\alpha}^{\times} & = & -F^{\alpha}_s \cos(2\psi) + F^{\alpha}_c \sin(2\psi)
\ena
where
\bea
F^{\alpha}_c & = & \left\{-
\frac1{4} (\sin^2(\chi_{\alpha}) - 2\cos^2(\chi_{\alpha}) ) \sin^2(\theta) - \right.\nonumber \\
& &  \left. \frac1{2} \cos(\chi_{\alpha}) 
\sin(\chi_{\alpha}) \sin(2\theta) \cos(\phi - \gamma_{\alpha}) + \right. \nonumber \\
& &  \left.   \frac1{4}(1 + \cos^2(\theta))\sin^2(\chi_{\alpha})\cos(2\phi - 2\gamma_{\alpha}) 
\right\} \frac{1}{1 + \hat{n}^{\alpha}\cdot\hat{k}}\nonumber\\
F^{\alpha}_s & = & \left\{ 
\cos(\chi_{\alpha})\sin(\chi_{\alpha})\sin(\theta)\sin(\phi - \gamma_{\alpha}) + \right. \nonumber \\
& &  \left. \frac1{2}\sin^2(\chi_{\alpha}) \cos(\theta)\sin(2\gamma_{\alpha} - 2\phi)
\right\}\frac{1}{1 + \hat{n}^{\alpha}\cdot\hat{k}}.
\ena
Note that $F^{\alpha}_c$ and $F^{\alpha}_s$ depend only on the sky locations of the pulsar $\{\gamma_{\alpha}, \chi_{\alpha}\}$ 
and of the GW source $\{\phi, \theta\}$. Using these expressions we can rewrite the residuals as 

\bea
\label{eq:residualsum}
r_{\alpha}^E(t) = \sum_{j=1}^4 a_{(j)} h_{(j)}^{\alpha},
\ena 
where
\bea
h_{(1)} = F^{\alpha}_c \sin[\Phi(t)], \;\;\; h_{(2)} =  F^{\alpha}_s \sin[\Phi(t)],\nonumber \\
h_{(3)} = F^{\alpha}_c \cos[\Phi(t)], \;\;\; h_{(4)} =  F^{\alpha}_s \cos[\Phi(t)],
\ena
and
\begin{eqnarray*}
a_{(1)} = \frac{\mathcal{A}}{2\pi f} [(1 + \cos^2{\iota}) \cos(2\psi) \cos(\Phi_0) - 2\cos{\iota} \sin(2\psi) \sin(\Phi_0)],\\
a_{(2)} = \frac{\mathcal{A}}{2\pi f} [(1 + \cos^2{\iota}) \sin(2\psi) \cos(\Phi_0) + 2\cos{\iota} \cos(2\psi) \sin(\Phi_0)],\\
a_{(3)} = \frac{\mathcal{A}}{2\pi f} [(1 + \cos^2{\iota}) \cos(2\psi) \sin(\Phi_0) + 2\cos{\iota} \sin(2\psi) \cos(\Phi_0)],\\
a_{(4)} = \frac{\mathcal{A}}{2\pi f} [(1 + \cos^2{\iota}) \sin(2\psi) \sin(\Phi_0) - 2\cos{\iota} \cos(2\psi) \cos(\Phi_0)].
\end{eqnarray*}

\subsection{Detection strategy}
We have now brought the timing residuals into a form that allows analytic maximization 
of the likelihood (to be introduced in the next paragraph) over the parameters $a_{(i)}$, in a way 
similar to the $F$-statistic. We introduce the inner product:
\bea
<x|h> \equiv \frac{T_o}{S(f)} (x || h), \;\;\; (x || h) \equiv \frac2{T_o} \int_0^{T_o} x(t) h(t) dt,
\ena
where $T_o$ is the observation timespan and $S(f)$ is the two-sided noise spectral density. 

Consider now a dataset $x_{\alpha}$; the log-likelihood that this dataset 
contains a GW signal $r_{\alpha}(t; \vec{\lambda})$~\footnote{Hereafter we omit the superscript ``E'', implicitly 
assuming that we are considering the Earth-term only.}, where $\vec{\lambda}$ is a vector of GW parameters 
describing the signal, is 
\begin{eqnarray}
\label{eq:likelihooddef}
\log{\Lambda_{\alpha}} \sim <x_{\alpha} | r_{\alpha}> - \frac1{2} <r_{\alpha} | r_{\alpha}> \sim 
 (x_{\alpha} || r_{\alpha}) - \frac1{2} (r_{\alpha} || r_{\alpha}).
\end{eqnarray}
Since we consider monochromatic sources and noise-less data, $S(f)$ is just a scaling constant in our
calculations; in the following, we neglect numerical coefficients, and we compute scaled log-likelihoods. 
Let us now assume we have a collection of datasets 
$x_{\alpha}$, $\alpha=1,...,P$, corresponding to observations of $P$-pulsars, and let us infer 
that a single GW is present in the datasets. The total log-likelihood evaluated on the $P$ datasets
is simply the sum of the individual log-likelihoods, and substituting $r_{\alpha}$ given
in equation (\ref{eq:residualsum}) into equation (\ref{eq:likelihooddef}), can be written as: 
\begin{eqnarray}
\label{eq:sumlikelihood}
\log(\Lambda) & = & \sum_{\alpha = 1}^P \log (\Lambda_{\alpha})\nonumber \\
& \sim & \sum_{j=1}^4 a_{(j)} X_{j} - 
\frac1{2} \sum_{j=1}^4 \sum_{k=1}^4 a_{(j)} a_{(k)} M_{jk}, 
\end{eqnarray}
where
\bea
\label{eq:XM}
X_{j} \equiv \sum_{\alpha=1}^P (x_{\alpha} || h^{\alpha}_{(j)}),\;\;\;\;   M_{jk} \equiv \sum_{\alpha=1}^P 
(h_{(j)}^{\alpha} || h_{(k)}^{\alpha})
\ena
As in the $F$-statistics case, we can maximize the log-likelihood over the extrinsic parameters $a_{(j)}$:
\bea
\frac{\partial \log(\Lambda)}{\partial a_{(j)}} = 0,\;\;\;\; \to a_{(k)} = M^{-1}_{kj}X_j, 
\label{E:MLE}\\
\{ \log(\Lambda) \}_{{\rm max}\{a_{(j)}\}} = \frac1{2} X_k M^{-1}_{jk} X_j
\label{E:fstat}
\ena
where the sum over the repeated indices $j,k=1,...,4$ is assumed. Note that such maximization
absorbs the four parameters ${\cal A}$, $\iota$, $\psi$ and $\Phi_0$; the maximized log-likelihood 
has therefore a form similar to the $F$-statistic and (for a fixed frequency) is function of the 
sky location $\{\phi, \theta\}$ of the GW source only. From now on, when we talk about log-likelihood, 
we mean log-likelihood maximized over $a_{(j)}$ according to equations~(\ref{E:MLE}) and (\ref{E:fstat}), and 
for an individual GW source is a function of two parameters only. The $M$-matrix is block-diagonal, and
is given by

\bea
M_{jk} = \sum_{\alpha} \left( \begin{array}{cccc} 
(F^{\alpha}_c)^2 & F^{\alpha}_c F_{\alpha}^s & 0 & 0 \\
F^{\alpha}_cF^{\alpha}_s & (F^{\alpha}_s)^2 & 0 & 0  \\
0 & 0 & (F^{\alpha}_c)^2 & F^{\alpha}_cF^{\alpha}_s \\
0 & 0 & F^{\alpha}_cF^{\alpha}_s & (F^{\alpha}_s)^2 
\end{array}\right).
\ena

Clearly one does not know a priori how many GW sources contribute to the dataset; the extension of the model 
to $N$ sources is mathematically straightforward. For each source we simply have four $a_{(j)}$ coefficients,
meaning that all the summations in equations (\ref{eq:sumlikelihood}) and (\ref{eq:XM}) have to be performed
over $j,k=1,...,4N$. Now the coefficients $a_{(j)}$ become a $4\times N$ array, 
$X_j$ is also a $4 \times N$ array, while the $M$-matrix is now a $4N \times 4N$,
2-D matrix
\bea
M_{jk} = \sum_{\alpha=1}^P (h_{(j)}^{\alpha} || h_{(k)}^{\alpha}).
\ena
(where indices $j, k$ run over $1, ..., 4N$) and it takes into account for correlations 
between the different GW signals. 

When considering a multi-source model we can also estimate the relative contribution of each individual 
source to the overall likelihood estimator. For each source the analytical maximization of the likelihood
according to equation (\ref{E:MLE}) provides an  estimate of the four parameters 
characterizing the waveform. The relative contribution of each source to the maximum likelihood solution
normalized to the strongest source can therefore be  written as
\be
\label{eq:cij}
C_{n} = \log(\Lambda) \frac{\sqrt{\sum_{j=1}^4{ (a_{(j)}^{n})^2}}}{{\rm max}_{m}\left(\sqrt{\sum_{j=1}^4{ (a_{(j)}^{m}})^2  }\right)}
\en
where $n,m=1,...,N$. In the above expression ${\rm max}_{m}$ identifies the brightest source in the dataset,
and the sums are over the four coefficients of each $n$ source. Note that $C_{n}$ is a function 
of the source inclination and of the overall GW amplitude, and is proportional to $\sqrt{h_{+}^2 + h_{\times}^2}$, 
as introduced in equation~(\ref{eq:strain}). Note that in this definition, 
if we have identical GW sources but at different 
sky locations, then all the individual $C_n$ are the same and are equal to the total likelihood.

In the following sections we will perform a search of the GW source sky locations by 
maximizing the log-likelihood given by equation (\ref{eq:sumlikelihood}).
We will compute the log-likelihood assuming a variable number of GW sources contributing to the 
residuals, to check whether our analysis allows us to infer the correct number of sources present in the 
dataset. As a result the log-likelihood parameter space is generally a function of $2\times N$
parameters, corresponding to the sky locations of the $N$ GW sources. In plotting multi-source sky maps 
we will usually show regions of the parameter space which are within 10~\% of the maximum likelihood, 
and we will call them ambiguity or error regions. Actual ambiguity regions depend on the SNR of 
the GW signals, and are therefore meaningless in the noise-less case, nevertheless we think that the 
introduced error regions are still useful to indicate the degeneracies/ambiguities in the parameter space. 

\section{Testing the algorithm performance}
\label{resolvability}

In this section we apply the theory outlined above to some test cases to assess its effectiveness in 
resolving individual sources (and dependence on the number of sources and on the number of pulsars 
in the array used for detection). We consider here the limit of few GW sources present in the dataset; 
extension of the analysis to a realistic population of sources is deferred to future investigations.  

\subsection{Minimum requirements for source localization}
\label{S:1-2GW}

We first test the statement that $3N$ pulsars are required to resolve $N$ GW sources. We start
by considering one GW source and one pulsar and we add pulsars and sources one by one to see how the 
likelihood pattern describing the source sky location changes. We start by injecting a first source at 
a location $\{4.26, 0.78\}$, then we add a second source at a location $\{1.50, 1.52\}$. Pulsars
are placed randomly in the sky.

\begin{figure}
\includegraphics[width=0.4\textheight, keepaspectratio=true]{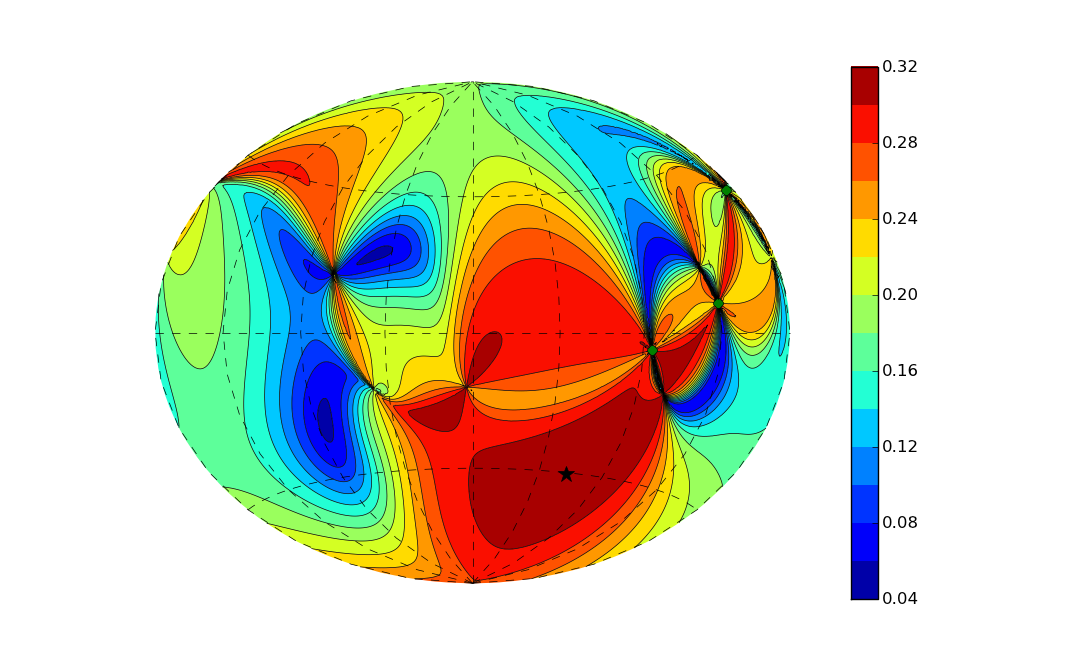}
\caption{Log-likelihood sky map for the sky  position of the GW source (black star), in a detection  
with three pulsars (green circles). Note that only relative values of the log-likelihood are meaningful. 
The best estimate for the source position is $\{4.27, 0.78\}$, whereas the true one is $\{4.26, 0.78\}$.}
\label{F:1GW3plsrs}
\end{figure}

\begin{itemize}

\item {\it One GW signal and one pulsar}. In this case we might only say whether the GW signal is 
present or not. Our method cannot estimate the parameters of the source. Mathematically this correspond
to a singular (determinant equal to zero) $M$-matrix. 

\item {\it One GW signals and two pulsars}.  With two pulsars we can estimate four parameters but we 
cannot get any information about the sky position: $\Lambda(\phi,\theta) =  const$, with the exact 
value depending on the relative position of the pulsars and the source. 

\item {\it One GW signal and three pulsars}. Finally, we have enough measurable quantities to estimate 
the position of the source on the sky. The accuracy depends on the relative position of the pulsars 
and the source. In figure \ref{F:1GW3plsrs} we give an example of the log-likelihood map
of the source sky position. The shape and 
area of the likelihood peak strongly depend on the relative position of the source and the pulsars. 
The presence of the noise will make the picture worse by (possibly) promoting false local maxima. Conversely, 
adding more pulsars significantly improves the accuracy of the localization. 

\begin{figure}
\centering
\begin{tabular}{c}
\includegraphics[width=0.40\textheight, keepaspectratio=true]{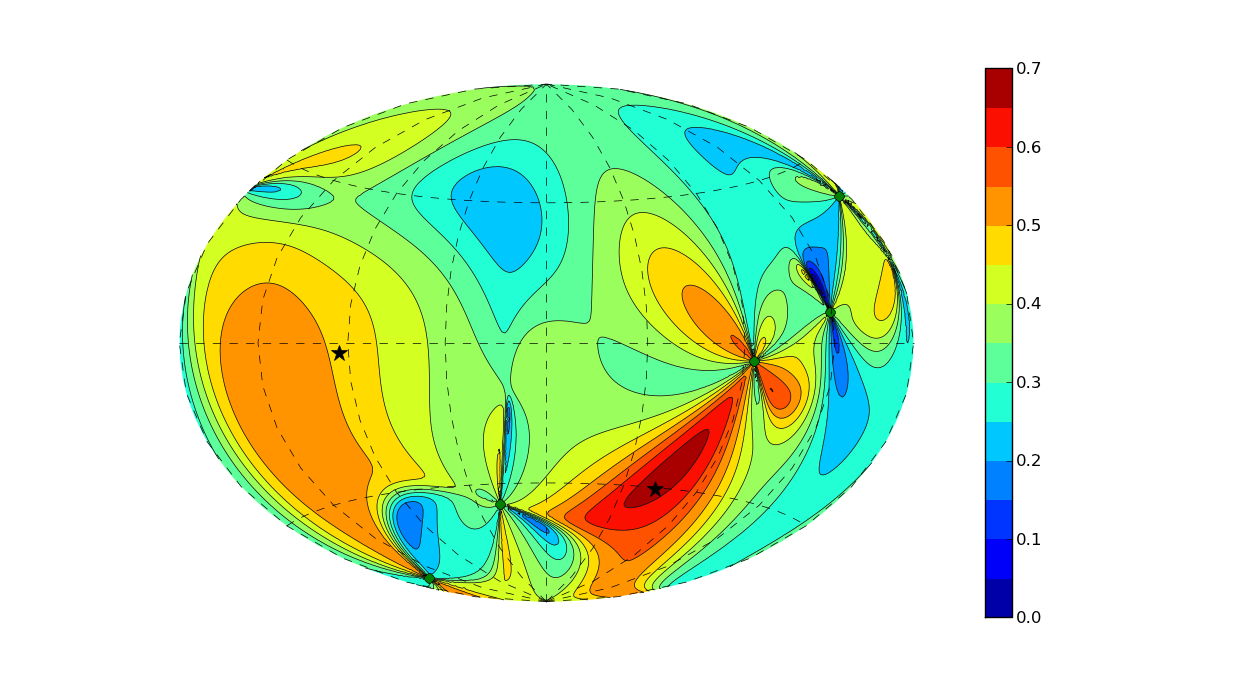}\\
\includegraphics[width=0.37\textheight, keepaspectratio=true]{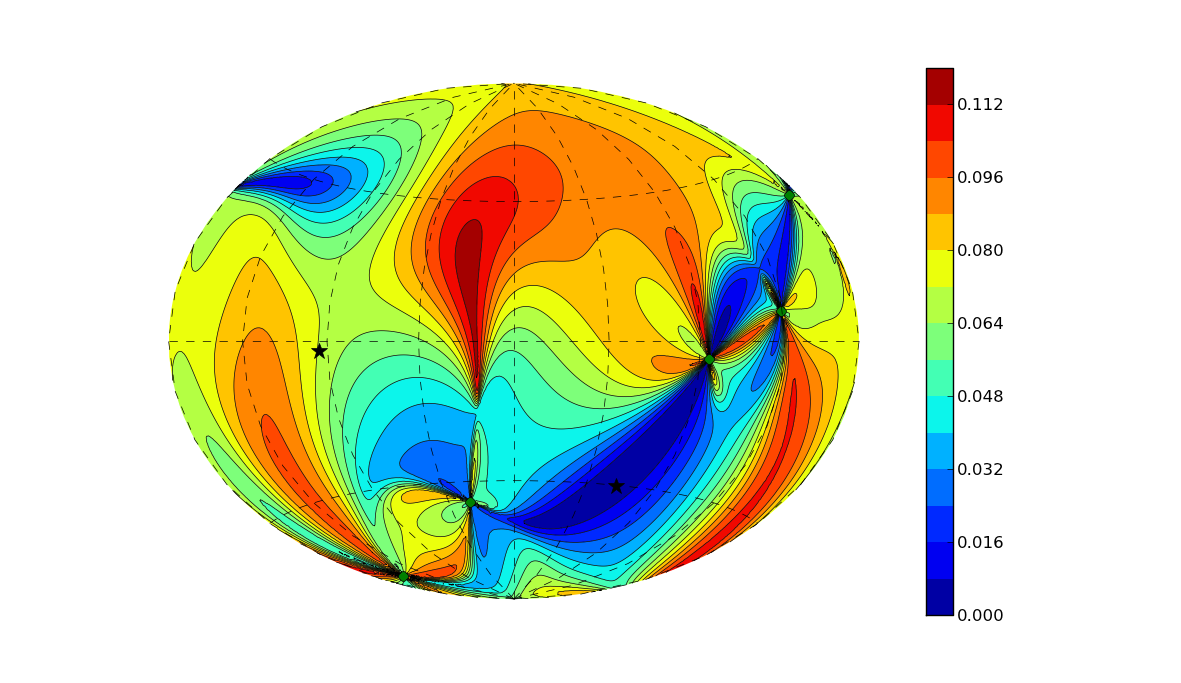}\\ 
\end{tabular}
\caption{Top panel: five pulsars (green circles) and two sources (black stars): the source 
log-likelihood map is constructed assuming a one source model, whereas there are two sources in the dataset. 
The best source position estimate  is $\{4.33, 0.89\}$; the true positions of the sources are 
$\{1.50,  1.52\}$ and $\{4.26, 0.78\}$. Bottom panel: log-likelihood map, assuming a
single GW source, constructed after the removal of the best estimate shown in the top panel.
Although there are hints of the presence of a second source, its inferred sky location 
does not match the actual position of the second GW source at $\{1.50,  1.52\}$.}
\label{F:2GW5plsrs}
\end{figure}

\item {\it Two GW signals and three pulsars}. We do not have enough measured quantities 
to determine the parameters of \emph{both} GW sources; again, 
the $M$-matrix is degenerate. However we can compute the likelihood map assuming that there is only 
one GW source contributing to the measured timing residuals. The result depends on the strength 
of the GW signals (both on the intrinsic strength and on the pulsar response function, i.e., 
on the relative source-pulsar sky location). We tested the most degenerate case: we took two
identical GW sources, and we placed them at different sky locations ($\{4.26, 0.78\}$, $\{1.50, 1.52\}$). 
Not surprisingly, being the signal a superposition of two monochromatic plane waves,
the two sources show up like a single one at a midway sky location. Adding a forth pulsar 
does not significantly improve the situation.


\item {\it Two GW signals and five pulsars.} Still we do not have enough measured quantities
(10) to measure all the source parameters (12). However, in a single GW source model, we 
start to see evidence of two likelihood maxima for the source sky location, a strong indication
that there is more than one source. As shown in the upper panel of figure~\ref{F:2GW5plsrs},
we now resolve one source and we have a large local maximum next to the second one. 
The result varies with the relative source-pulsar position, but for fixed (chosen above) 
positions of the GW sources we usually resolve one source or the other. 
In addition we have computed the log-likelihood map assuming two GW sources in the data. 
Since we still do not have enough measurements to estimate all the parameters, the likelihood forms 
a complicated pattern in the sky reflecting this degeneracy. However the value of the log-likelihood 
for the two-source model is higher (by about 13\%) than the single source model, strongly favoring
the presence of two signals in the data. To further check what can be possibly done in this situation,
we have tried to take the best estimate obtained by the single-source model, subtract it from the 
dataset and re-analyze it again searching for the second GW source (using again a single-source model). 
Unfortunately the procedure does not work well. The maximum likelihood estimators of the amplitude 
parameters of the first source have large errors (estimated values are 
$a_{(i)} = \{6.15,  -1.71,  -5.44,  -2.57\} \times 10^{-10}$ 
instead of the correct values: $a^{true}_{(i)} = \{ 4.08,  -2.54,    -2.53,    -4.09\}\times 10^{-10}$): 
when subtracting it, we are effectively partially removing the second GW source. The result after the
removal is shown in the lower panel of figure~\ref{F:2GW5plsrs}. There is an indication of a 
second GW signal but it is in a completely wrong location. We can conclude that in an array of 
five pulsars there is enough information to identify the presence of two sources, but it is 
impossible to correctly localize {\it both} of them in the sky.

\begin{figure}
\includegraphics[width=0.4\textheight, keepaspectratio=true]{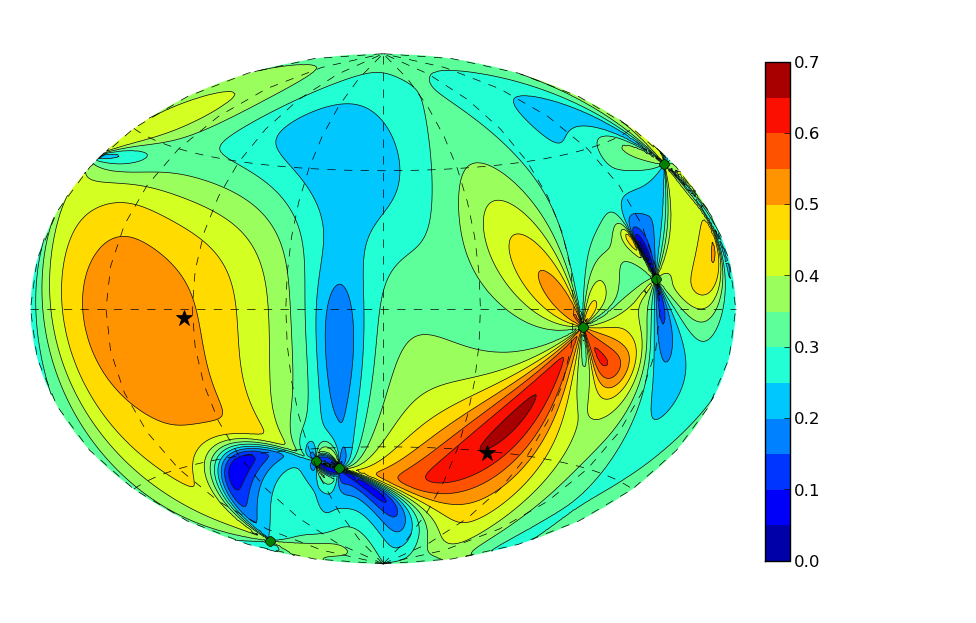}
\caption{Same as the top panel of figure \ref{F:2GW5plsrs}, but now using six pulsars for the 
detection. The log-likelihood map clearly shows two extended maxima around the true location of the 
sources. The best single source sky location estimate is $\{4.35, 0.94\}$,  the true positions
are the same as in figure \ref{F:2GW5plsrs}.}
\label{F:2GW6plsrs}
\end{figure}

\item {\it Two GW signals and six pulsars.}
Finally we have enough pulsars to estimate all the GW parameters, however the randomly chosen sixth pulsar is 
quite close to one of the previous five, so some level of degeneracy remains. The log-likelihood contour plot
for the single source model shown in figure \ref{F:2GW6plsrs} clearly has two maxima centered at the true 
location of the sources. We also successfully tested a two source model: the pair of points in the sky which 
returned the highest log-likelihood were $\{1.48,  1.55\}, \{4.31, 0.78\}$, close to the true values. 

\begin{figure}
\centering
\begin{tabular}{c}
\includegraphics[width=0.40\textheight, keepaspectratio=true]{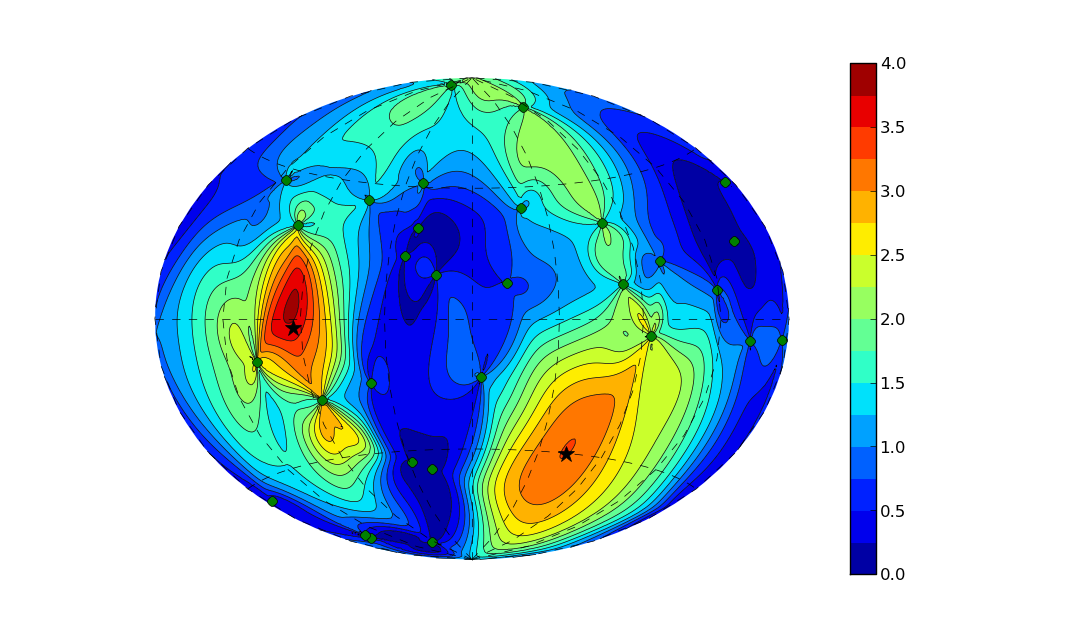}\\
\includegraphics[width=0.30\textheight, keepaspectratio=true]{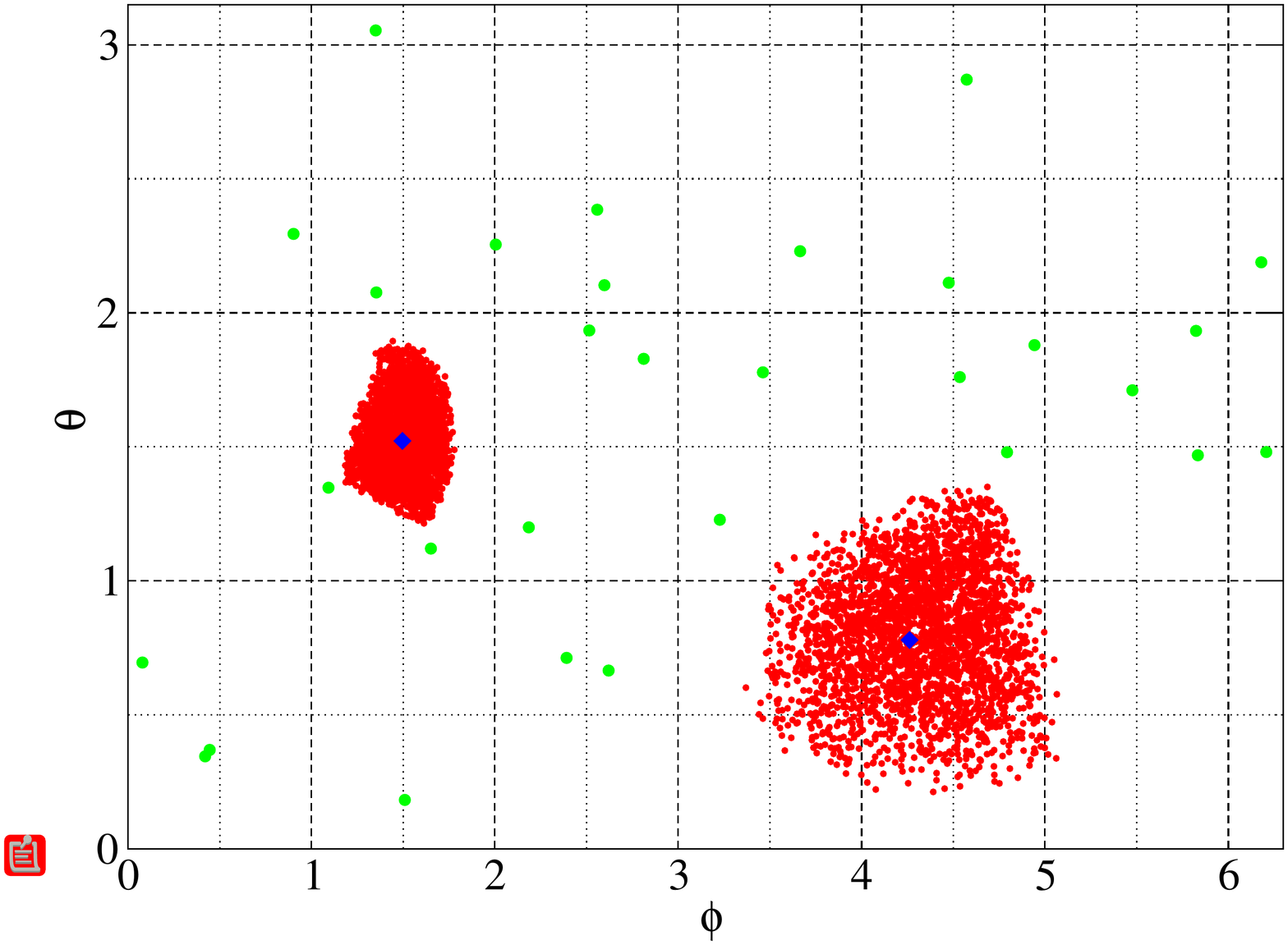}\\ 
\end{tabular}
\caption{Top panel: same as in figure \ref{F:2GW5plsrs}, but now using 30 pulsars for 
the detection. The best single source position estimate is now shifted to the other 
source, and the inferred location is $\{1.48, 1.63\}$, whereas the true position is 
$\{1.50, 1.52\}$. Bottom panel: log-likelihood map assuming that there are two  GW 
sources (blue diamonds), with measurements from 30 pulsars (green circles). We have 
selected all the pairs of points which are within 10\% of the maximum likelihood estimate. 
The best source sky location estimates are $\{1.48,  1.55\}, \{4.31, 0.78\}$, whereas 
the true positions are $\{1.50, 1.52\}, \{4.26, 0.78\}$.}
\label{F:2GW30plsrs}
\end{figure}

\item {\it Two GW signals and thirty pulsars.} By adding pulsars to the array, uncertainty regions shrink 
and secondary maxima disappear. Here is an example of what we would see with data from 30 pulsars. 
The log-likelihood contour plot for the single source model shown in the top panel of figure 
\ref{F:2GW30plsrs} nails down the true location of the sources. We plot the results of a two 
source model in the lower panel of figure \ref{F:2GW30plsrs}.
In this case, the likelihood is a function of four parameters (the angles defining the sky location of 
the two sources) and a visual representation is not straightforward. We randomly choose pair of points 
in the sky and computed the log-likelihood. In the figure we plot all the pair of points which are 
within 10\% of the maximum likelihood estimate (note that a 10\% drop in the log-likelihood roughly 
corresponds to the uncertainty region of a pair of sources with combined SNR$=10$).
\end{itemize}

\subsection{Resolving a cluster of GW sources}

\begin{figure}
\includegraphics[width=0.4\textheight, keepaspectratio=true]{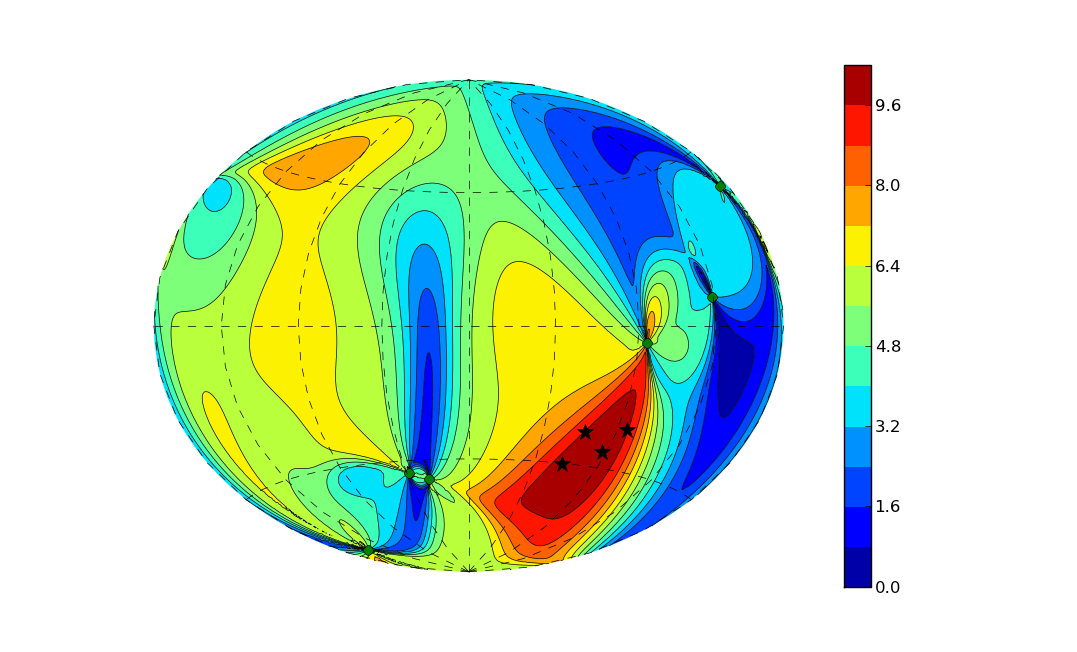}
\includegraphics[width=0.4\textheight, keepaspectratio=true]{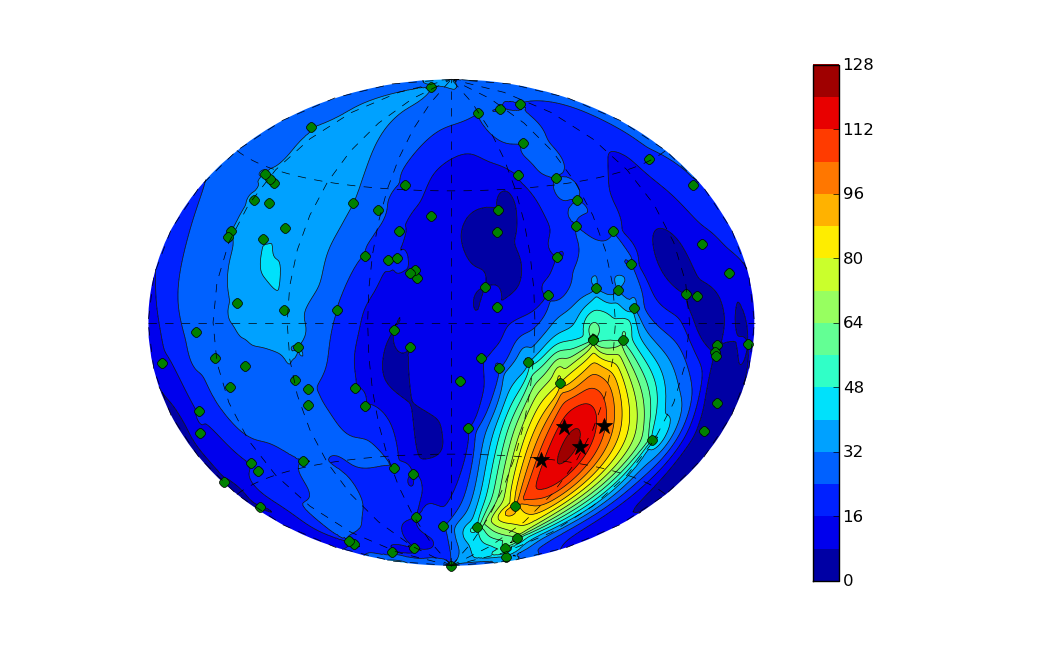}
\caption{Detection of a cluster of sources (black stars in each plot). The log-likelihood sky map for a 
single source model is shown, assuming detection with 6 (top) and 100 (bottom) pulsars (green circles).
The color coding corresponds to the (scaled) log-likelihood value. 
}
\label{F:Clust_single}
\end{figure}

In the previous subsection we looked at the detection and sky location of one or two GW sources with 
variable number of pulsars. We have chosen on purpose widely separated sources. We consider now a small 
group (cluster) of GW sources located close to each other in the sky. The idea is to check whether we can still
infer the number of sources in the cluster and resolve them individually. We consider sources of equal 
intrinsic strength. We start with computing the log-likelihood sky map assuming a single GW source and 
incrementing the number of pulsars from 6 to 100. Even with 100 pulsars the cluster looks like a single 
maximum in the likelihood extending over the whole cluster. The size of this maximum shrinks as we 
include more and more pulsars, but we do not see the development of multiple maxima corresponding 
to the true locations of the sources. We show likelihood contour plots in figure~\ref{F:Clust_single} 
for 6 and 100 pulsars in the array.

We then tried multiple GW source models. We started with 6 pulsars; in this case we cannot try models with 
more than two GW sources. The best guess for a two-source model is $\{4.60, 0.90\}; \{4.22,  0.96\}$, to 
be compared to the true positions of the sources:  $\{4.26,  0.78 \};  \{4.37,    0.98\};  \{4.66,  0.88\};  
\{4.82,    1.01\}$. Increasing the number of pulsars to 12, we can test models with up to 4 sources.
Unfortunately the result is a significant increase in the ambiguity area with poor source localization.
For example, the best guess with a three source model is 
$\{4.84,  2.03,  0.0004\};   \{4.28,  0.88,  0.123\}; \{4.71,  0.94,  0.161\}$. 
The third number in each parenthesis is the estimate of the relative contribution of the 
source to the log-likelihood, i.e., $C_{n}$ as defined by equation (\ref{eq:cij}). The first source in the
solution contributes negligibly to the result: the best solution is essentially still a two-source model. 

The result improves by increasing the number of pulsars and running models with 2-, 3- and 4-sources. 
The best guess with 100 pulsars is close to the true positions of the GW sources, and all four 
points give a significant contribution to the likelihood: 
$\{4.20,   0.81,   0.67\}$;  $\{4.41, 0.92,  1.19\}$; $\{4.63, 0.89,  1.2281\}$; $\{4.90, 1.04, 0.65\}$. 
However the maximum likelihoods of the best 1-, 2-, 3- and 4-source 
solutions are very similar: 1.2144, 1.2279, 1.2281, 1.22815 
(to be compared to a true likelihood of 1.22818). This means that we might not be able to 
differentiate between those models in practice (in presence of noise).  

\subsection{Determining the number of GW sources}
\label{nsources}

\begin{figure}
\includegraphics[height=0.25\textheight, keepaspectratio=true]{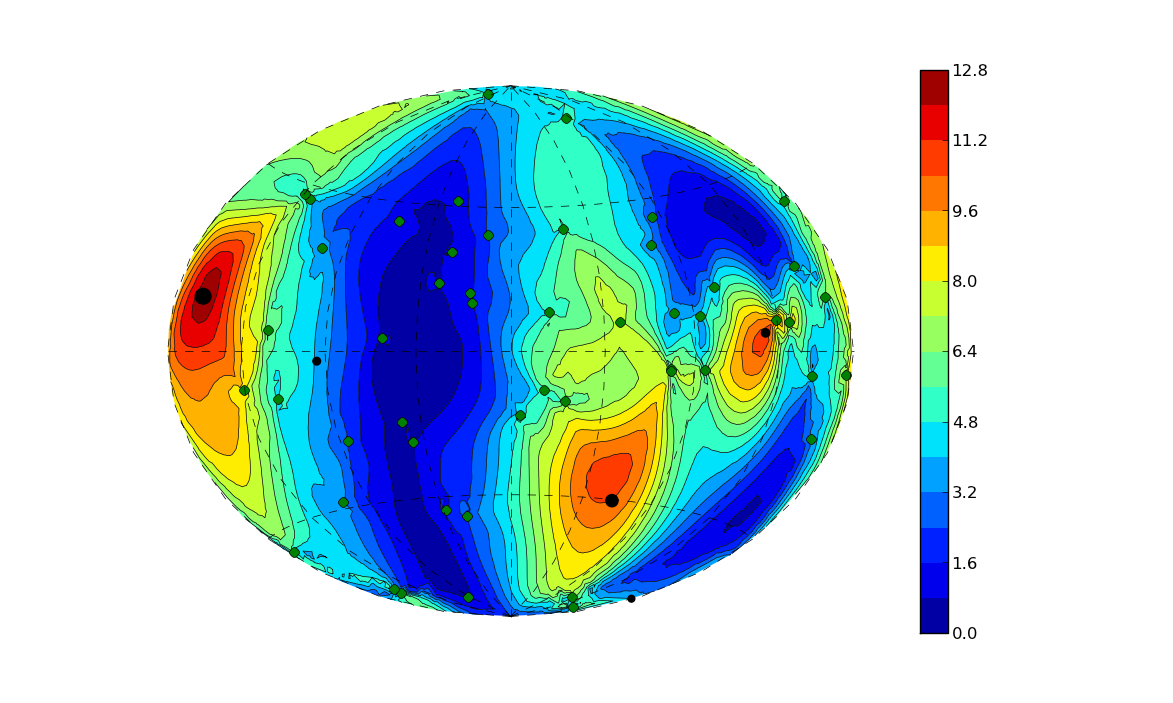}
\caption{Log-likelihood 
sky map for the single-source model in the case of five unequal GW sources detected with an array of 50 pulsars. 
The true position of the five GW sources is shown as black circles, with relative size corresponding 
to their relative strength. Pulsars are shown as green dots.}
\label{F:Scat_single_r}
\end{figure} 
We consider now a signal generated by the superposition of five sources with known sky location and relative
strength. We test two cases: in the first case, all sources have equal strength; in the second one, we take
the relative source strengths of the five brightest sources found in a realization of the cosmic SMBHB population
from \cite{KS11}. Sources are placed randomly in the sky, and their location is the same for both cases.

As a first test, we estimate the likelihood function of a single-source model as a function of the number $P$ of pulsars
in the array. Results are similar in the two cases. In the equal strength case, when $P$ is small, we observe a single maximum
resulting from the combination of the five sources. When $P\approx30$,  we detect two local maxima corresponding 
to the true location of two of the GW sources, and when $P\approx50$, the local maxima correctly recover three 
sources. A similar situation occurs in the unequal case, with the difference that now, when $P$ is small, the 
single maximum corresponds to the actual position of the brightest source. Again, increasing $P$ up to 50, we 
observe three maxima, corresponding to the three brightest sources. The log-likelihood sky map
for $P=50$ in the unequal case is shown in figure~\ref{F:Scat_single_r}.

\begin{figure*}
\centering
\begin{tabular}{cc}
\includegraphics[height=0.25\textheight, keepaspectratio=true]{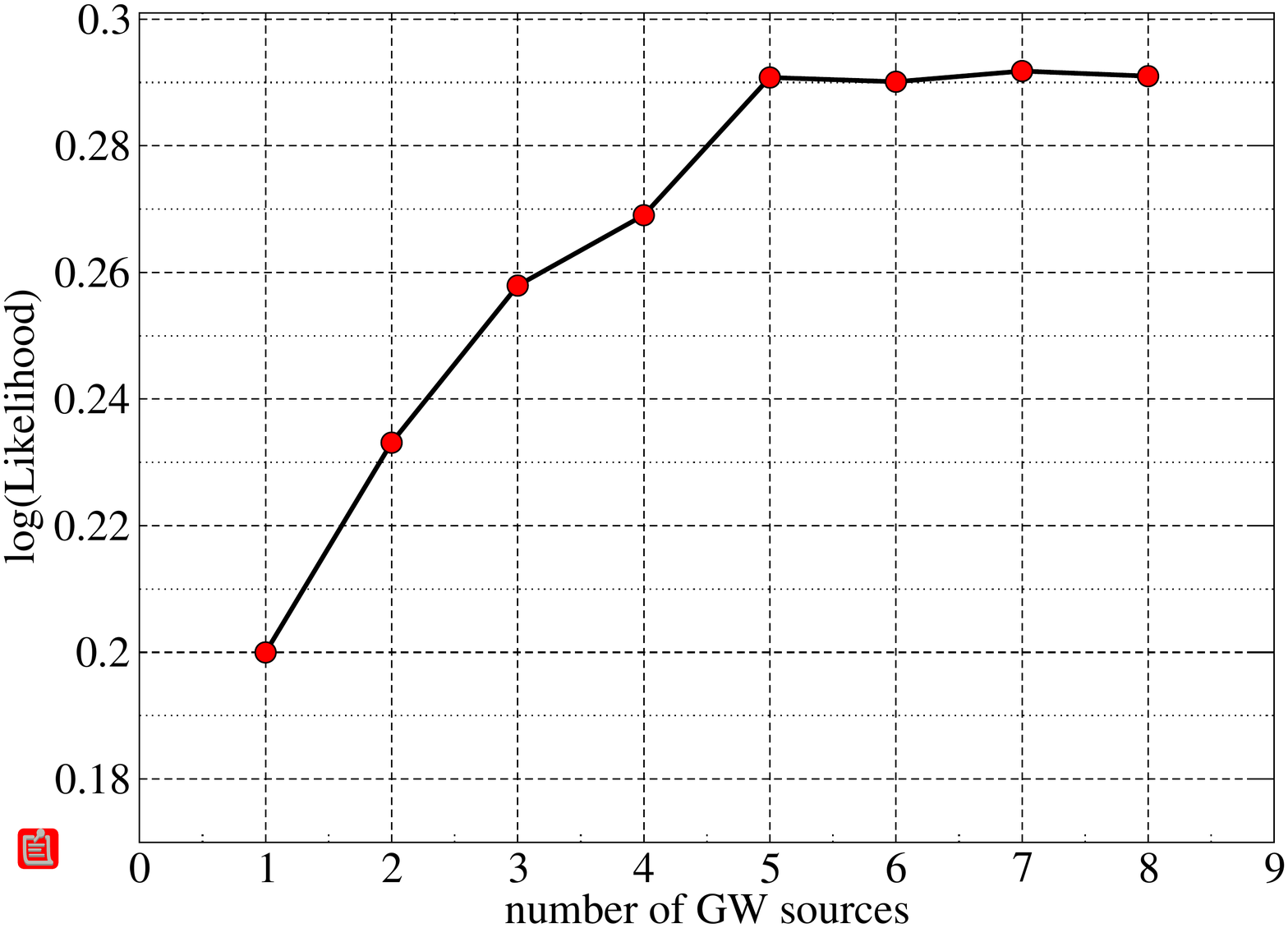} &
\includegraphics[height=0.25\textheight, keepaspectratio=true]{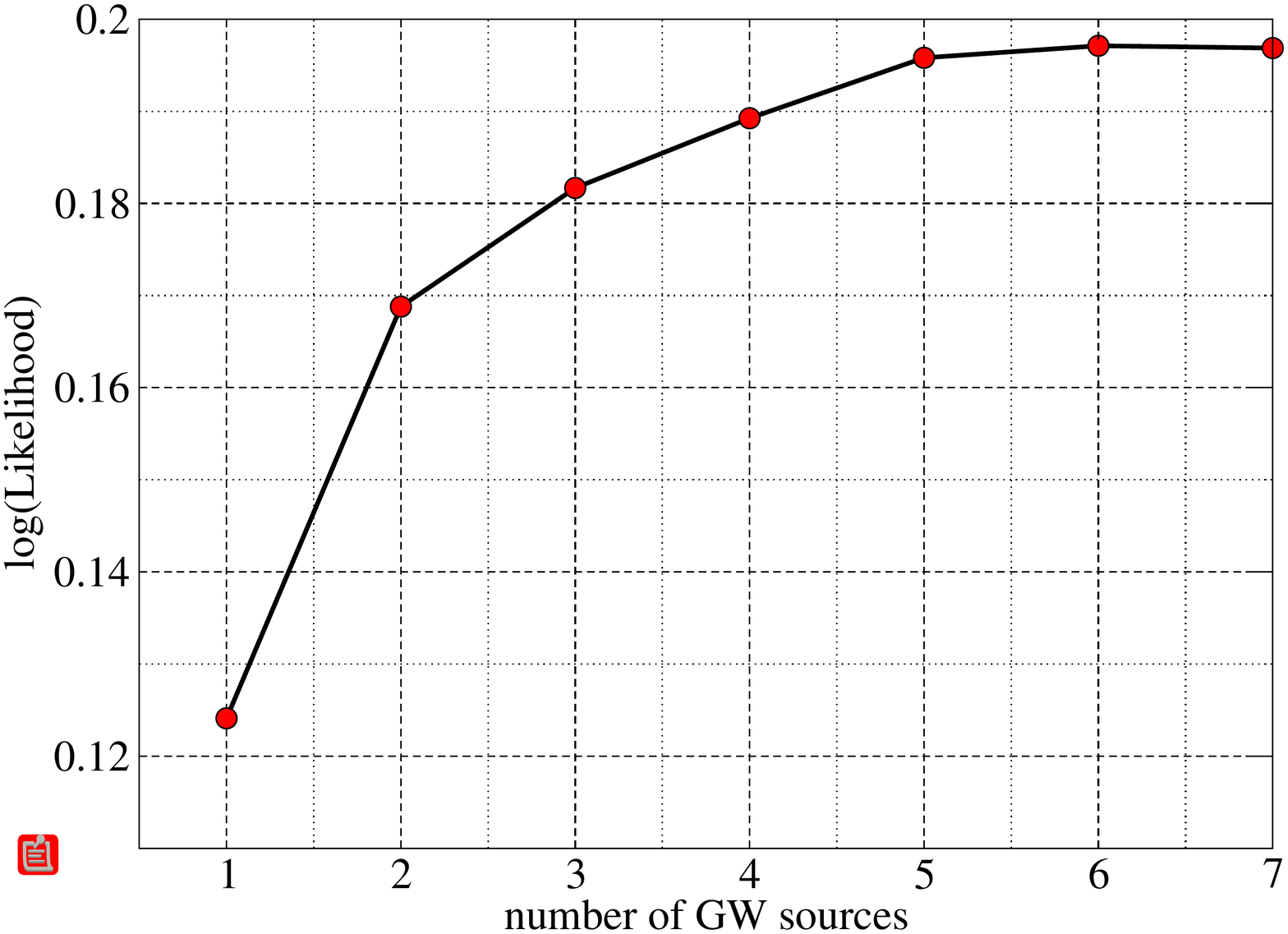}
\end{tabular}
\caption{Maximum estimated log-likelihood of the $N$-source models as a function of $N$.
The dataset contained five GW sources, and detection was performed by
an array of 50 pulsars. Left panel: equal strength sources; right panel: unequal strength sources.
In both cases, the maximum log-likelihood flattens out for models with more than five sources.  
}
\label{F:L_vs_n}
\end{figure*} 

Using 50 pulsars we then tried 2-, 3-, 4-, 5-, 6-, 7-, 8-source models.  We should emphasize here that our
maximum likelihood search is far from being optimal, so we had to do quite extended searches for models with 
more than 4 GW sources. We plan to apply proper search algorithms to this problem in future work.
In figure~\ref{F:L_vs_n} we show the maximum recovered log-likelihood as a function of the number of GW 
sources included in the model for the two examined cases. In the equal case, we see a steady increase 
of the maximum likelihood up to five sources, beyond which the curve suddenly flattens out.
A similar behaviour is seen in the unequal case, with the difference that now the maximum likelihood
growth for 3-to-5 source models is much shallower than before. This is simply due to the unequal strength
of the sources. Also in this case, the likelihood flattens out if more than five sources are included. We 
can therefore safely say that, at least in the limit of few sources contributing to the signal, the algorithm 
is capable of unambiguously identifying the number of sources present in the data. 

\begin{figure}
\centering
\begin{tabular}{cc}
\includegraphics[width=0.17\textheight, keepaspectratio=true]{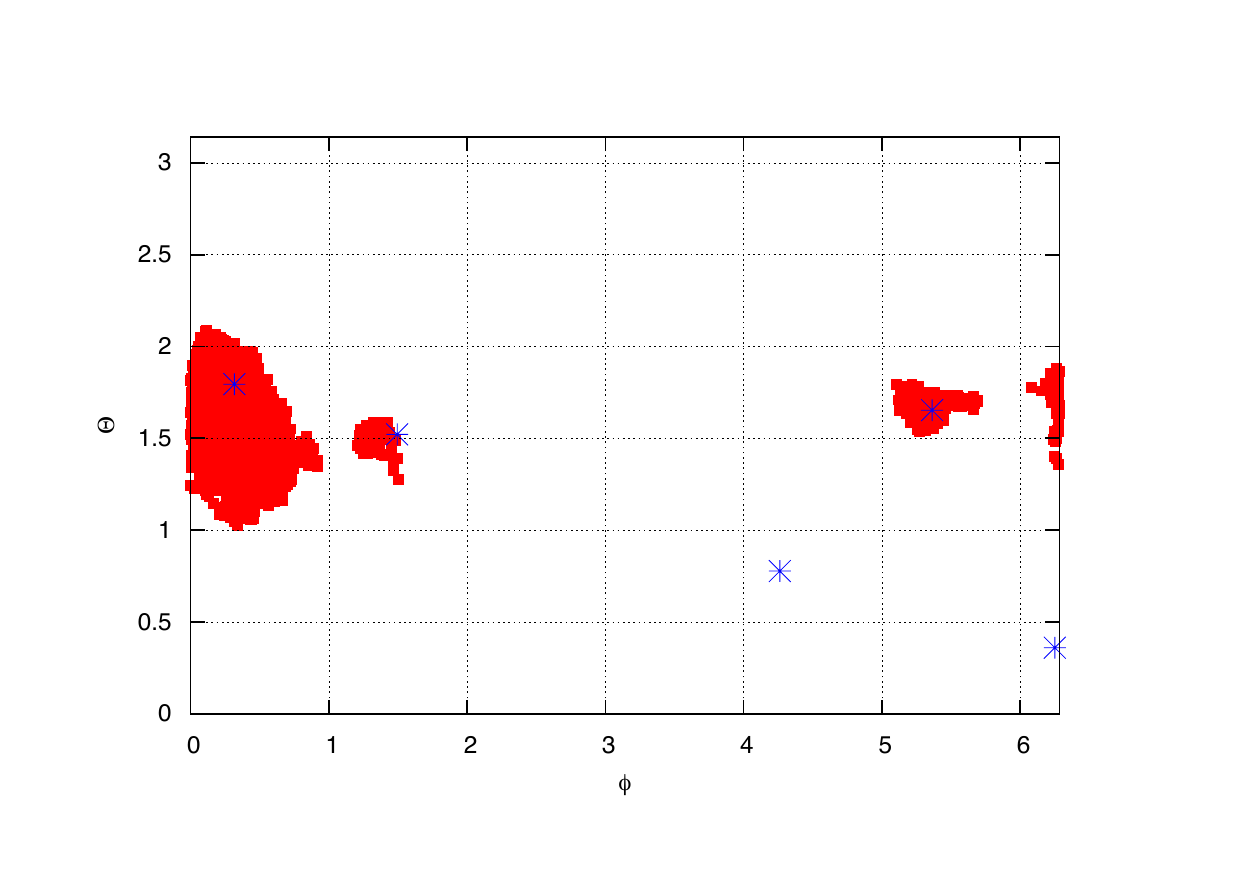} &
\includegraphics[width=0.17\textheight, keepaspectratio=true]{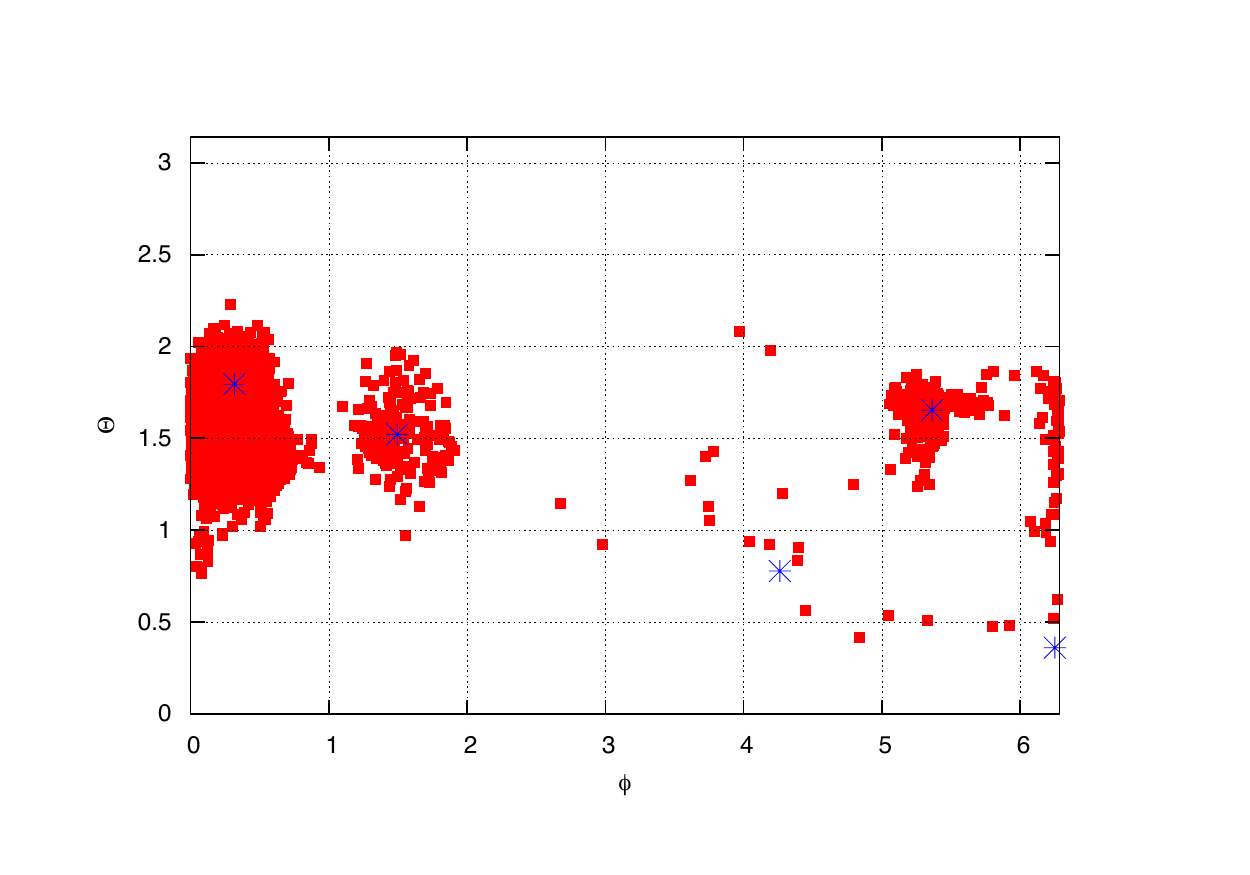}\\
\includegraphics[width=0.17\textheight, keepaspectratio=true]{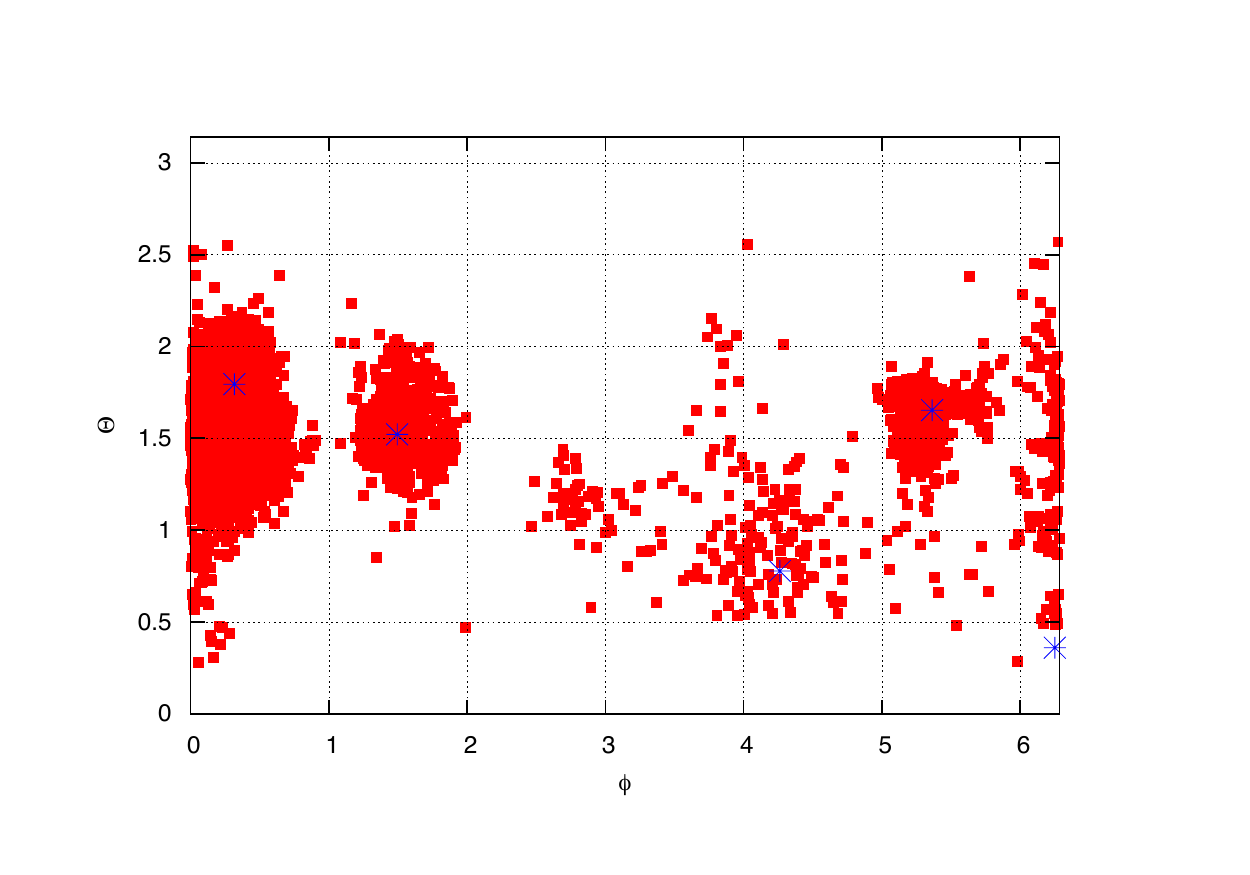} &
\includegraphics[width=0.17\textheight, keepaspectratio=true]{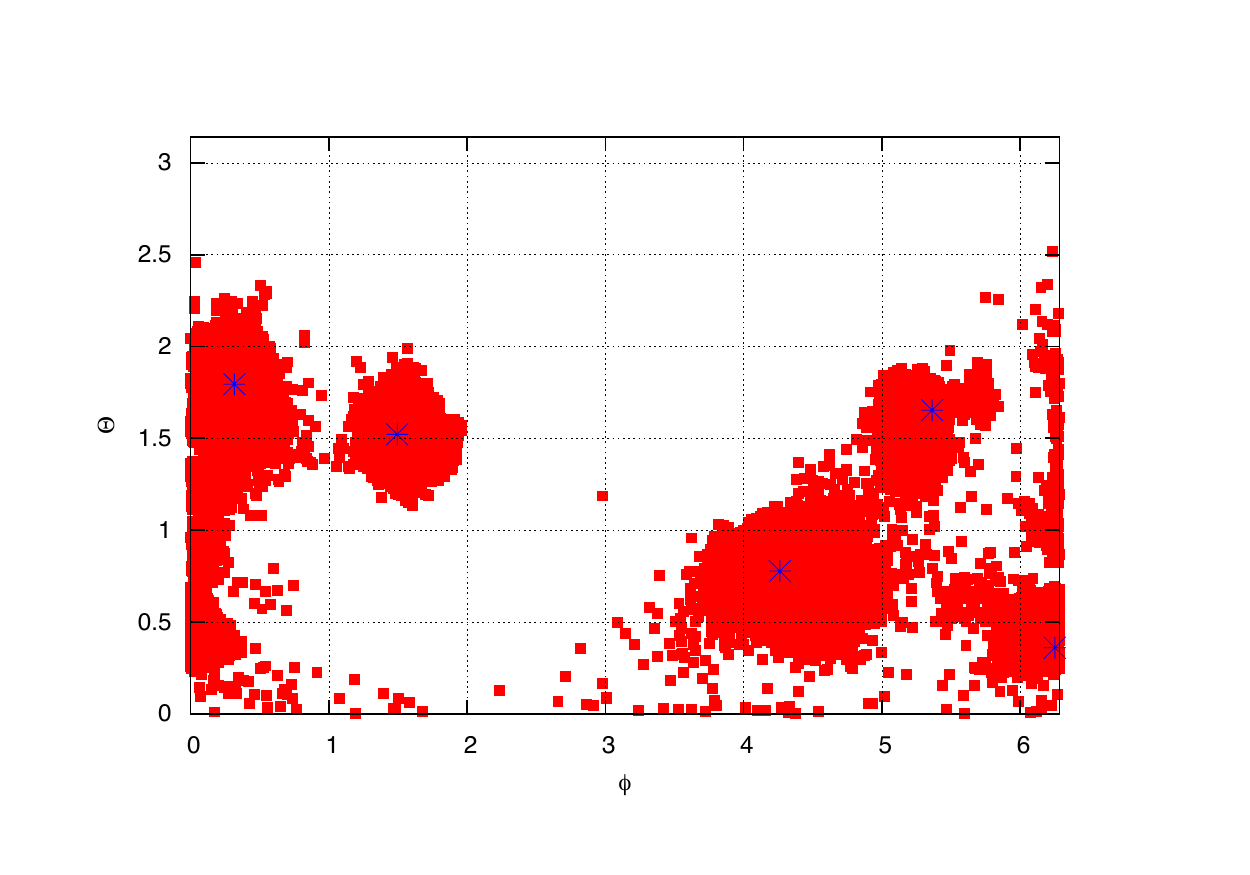}\\
\end{tabular}
\caption{Log-likelihood sky maps of the source sky positions for an injection of 5 source with equal strength,
detected with an array of 50 pulsars. From the top left to the bottom right, panels show the likelihood for 
a 2-, 3-, 4- and 5-source model. We plot all the points which have $C_{n}$ within 10\% of the 
maximum likelihood for each model.}
\label{F:Scat_50plsr}
\end{figure}

\begin{table*}
\centering
\caption {\label{T:Scat50Plsr_EqStr} Recovered sky position of GW sources for models with different number of sources.
For each model we also give the maximum likelihood estimator of the sky location $\{\phi, \theta, C_{n}\}$.}
\begin{tabular}{@{}ccccccc@{}}
\hline
\hline
Equal strength sources & & & & & &\\
\hline
True position & 2-source model & 3-source model & 4-source model & 5-source model & 6-source model & 7-source model\\
\hline
4.262,    0.779 &  -                  &  -                &  -                 & 4.24, 0.57, 0.29  & 4.43, 1.11, 0.22 & 4.40, 0.72,0.24\\ 
1.496,    1.522    & -                 & 1.43, 1.46, 0.21  &  1.46, 1.50, 0.27 & 1.49, 1.53, 0.23 & 1.52, 1.67, 0.25 &  1.50, 1.52, 0.25\\ 
5.363,    1.653 &  5.43, 1.69,  0.17 & 5.38, 1.68, 0.25 &  5.42, 1.68, 0.22 & 5.30, 1.63, 0.26  & 5.35, 1.62, 0.25 & 5.27, 1.63, 0.27\\ 
6.250,    0.361 &    -                 &      -            &  6.2, 1.03, 0.20 & 6.22, 0.32, 0.29 & 6.21, 0.42, 0.25  & 6.20, 0.38, 0.29\\ 
0.317,    1.794  &  0.57,  1.49,  0.23 & 0.28, 1.55, 0.26 & 0.37, 1.99, 0.21 &  0.45 1.81, 0.22 & 0.39, 1.86, 0.29 & 0.45, 1.73, 0.28 \\ 
-                 &  -                    & -                 &  -                 &  -                 & 1.08, 1.42, 0.04 & 1.36, 2.17, 0.04\\ 
-                 &  -                    & -                 &  -                 &  -                 &   -              & 2.92, 1.70, 0.03\\ 
-                 &  -                    & -                 &  -                 &  -                 &   -              &  -\\ 
\hline
\hline
Unequal strength sources & & & & & &\\
\hline
True position & 2-source model & 3-source model & 4-source model & 5-source model & 6-source model & 7-source model\\
\hline
4.262,    0.779,  0.83 &  4.34, 086, 0.13   &  4.22, 0.58, 0.17  &  4.37, 0.61, 0.19 & 4.42, 0.66, 0.20  & 4.39, 0.70, 0.17 & 4.37, 0.74, 0.16\\
1.496,    1.522,   0.52 & -                    & -                   &  1.62, 1.35, 0.08 & 1.61, 1.55, 0.11    & 1.45, 1.44, 0.10 &  1.49, 1.41, 0.10\\
5.363,    1.653,   0.58 &  -                   & 5.25, 1.60, 0.13  &  5.19, 1.70, 0.10  & 5.26, 1.68, 0.12   & 5.28, 1.65, 0.12 & 5.19, 1.61, 0.11\\
6.250,    0.361,   0.44 &    -                 &      -             &  -                  & 5.97, 0.39, 0.07  & 6.03, 0.33, 0.08  & 6.28, 0.30, 0.07\\
0.317,    1.794,   1.0   &  0.30,  1.73,  0.17 & 0.41, 1.68, 0.18  & 0.35, 1.65, 0.16   &  0.30 1.85, 0.19   & 0.30, 1.78, 0.20 & 0.34, 1.82, 0.20\\
-                 &  -                    & -                 &  -                 &  -                            & 5.09, 2.31, 0.01 & 3.04, 2.23, 0.01\\
-                 &  -                    & -                 &  -                 &  -                 &   -                           & 5.64, 2.74, 0.02\\
\hline
\end{tabular}
\end{table*}

In figure~\ref{F:Scat_50plsr} we  try to give a sense of how the likelihood map behaves for models with 
more than one source. As already pointed out before, a visualization of the likelihood is not straightforward, since 
it is now function of $2\times N$ parameters, and is not easily represented in a 2-D plot. 
Here we consider the equal strength case, and we show the points giving a contribution $C_{n}$ within 
10\% of maximum likelihood value. In the 2-source model (upper left panel) we already clearly see three clusters around
the true location of three of the injected sources. The situation does not significantly change adding a third source, 
but clusters around all the five sources begin to appear in the 4-source model (lower left panel) and are clearly
resolved in the 5-source model (lower right panel). If we go beyond a 5-source model, the likelihood map becomes 
more and more diluted (increase of the ambiguity regions) without showing any significant new ``cluster of points'',
or any improvement of the maximum likelihood. Note that also in the 5-source model, the ambiguity areas are quite 
large, which would, for example, pose a serious problem to a putative search for electromagnetic counterparts to the 
sources \cite{sesana11,tanaka11}. 
However we have noticed that the ambiguity regions around the bright sources in the 2-source 
model search (see upper left panel) are notably smaller than the areas around the corresponding sources
in the 5-source model search (lower right panel). We want to take this into account while constructing a proper
search algorithm, by restricting the ambiguity region in the five source search to that found in the 2-3 source 
model. This should allow us more accurate localization of the GW sources.


As a final step, we give the best recovered source parameters for each $N$-source model in 
table~\ref{T:Scat50Plsr_EqStr}. 
Notice that for a noiseless dataset, in principle, for a 5-source model, the best answer giving the maximum likelihood 
is the true one. However it might take quite a long time to converge to that answer. 
What we show in the table is the result of our (limited) search algorithm, which should give some ideas about the abilities to find and locate maxima.
The detection strategy performs well in both cases. As $N$ increases up to 5, we recover $N$ sources almost at the
right sky location. And for $N>5$, the additional sources always have a relative contribution to the likelihood which
is an order of magnitude smaller than the true ones. Notice also that the recovered values of $C_{n}$, representing
the relative contribution of the sources to the signal, are very close to the injected relative source strengths.
In the future work we intend to apply advanced search algorithms \cite[see, e.g.,][]{Littenberg:2009bm, 
vanderSluys:2008qx, Petiteau:2010zu, Feroz:2009de} developed and used 
 for GW data analysis in present and future interferometric detectors. 

\section{Blind searches}
\label{blind}

After testing the effectiveness of our detection strategy, we performed two sets of blind searches. 
For this exercise, datasets were generated by A. Sesana, and searches were performed by S. Babak without any 
previous knowledge of the content of the datasets, except for the fact that there were less than 10 sources.

\subsection{Noiseless data}
\begin{figure*}
\centering
\begin{tabular}{cc}
\includegraphics[width=0.35\textheight, keepaspectratio=true]{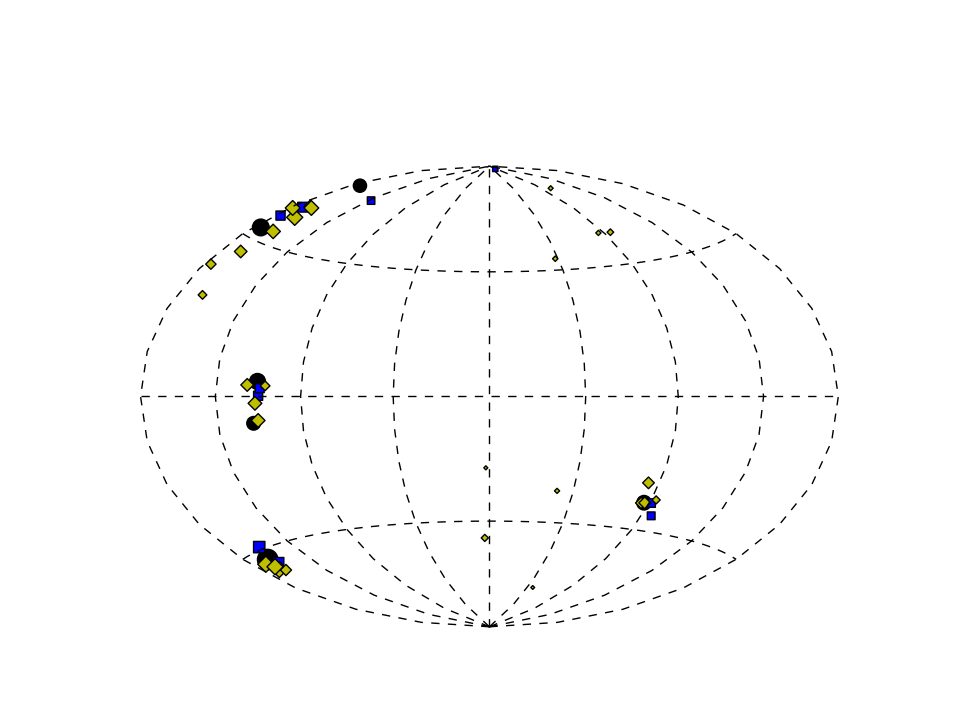} &
\includegraphics[width=0.35\textheight, keepaspectratio=true]{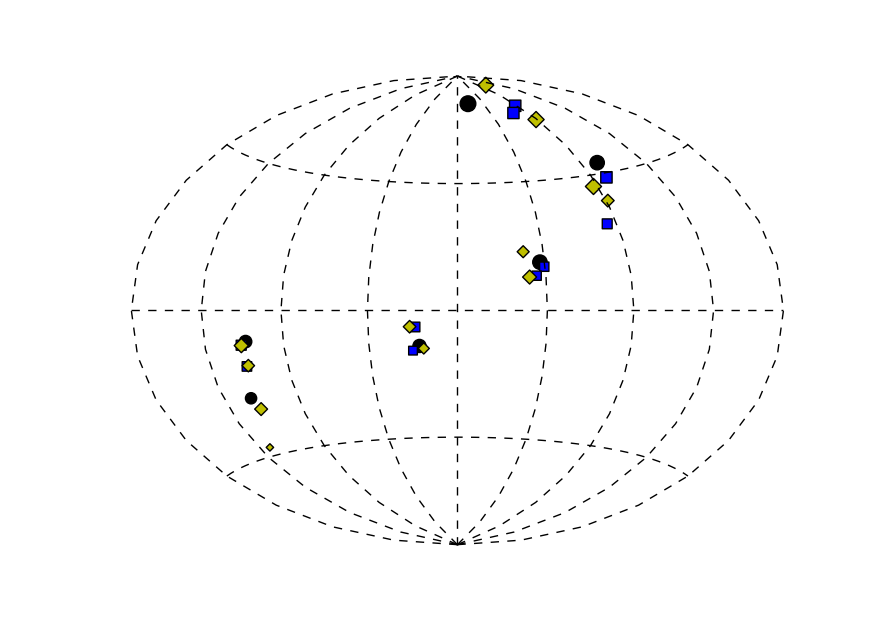}\\ 
\end{tabular}
\caption{Results of the noiseless blind searches. Left panel: few points from the best 5-source (blue) and 
7-source (yellow) solutions for the first dataset. Right plot: best guesses from the 5-source model (blue) and 
from the 6-source model (yellow). The relative size of the symbols indicates the relative contribution of each point to 
the combined likelihood. In both panels, the injected sources are shown as filled circles, with relative size
corresponding to their relative strength.}
\label{F:Blind}
\end{figure*}

In the first round of blind searches, we used noiseless datasets. We used equally sampled (once every two weeks)
time series for 50 pulsars located randomly in the sky. Two datasets were analyzed, both of them containing
six sources with unequal strengths at a frequency $f=2\times10^{-8}$Hz. The first search inferred 5 or 7 sources 
(corresponding to maximum likelihoods of 0.131 and 0.133, to be compared with  0.1345 which is the likelihood for the actual 
position of the 6 injected sources). In the second blind search the best guesses were 5 or 6 sources with maximum 
likelihood of 0.0699 and 0.071 respectively (to be compared to 0.077 for the true injection).
In the first dataset there were two close pairs of sources, while in the second one there was only 
one tight pair. The presence of tight pairs makes the source recovery problematic, especially because we do
not use any optimized search algorithm (note that in both cases, the maximum estimated likelihood
is quite smaller than the true one). Results are visualized in figure ~\ref{F:Blind}. 
Here, filled black circles denote the true location of the GW sources, and coloured diamonds and squares correspond 
to points extracted by the best model solutions. The relative size of the symbols represent either the relative 
strength of the sources, or the relative contribution to the likelihood of the recovered points.
Isolated sources are clearly identified, and their inferred position is usually within few degrees of the true
one. Close pairs are more problematic, because sometimes the search algorithm recognize a single source, 
midway between the two. We expect these results to  improve in a optimized search.

\subsection{Noisy data}
\begin{figure}
\includegraphics[height=0.3\textheight, keepaspectratio=true]{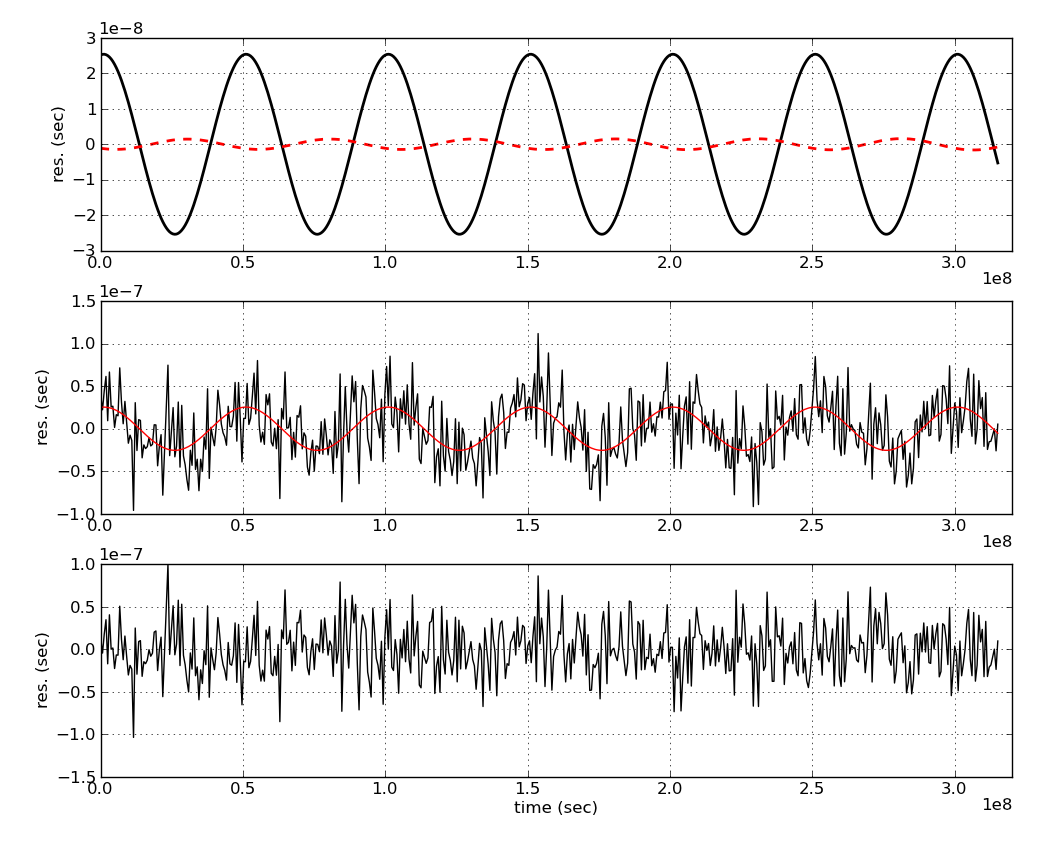}
\caption{Example of residuals used in the blind search. The dataset contained a superposition
of five monochromatic sources at $f=2\times10^{-8}$Hz, shown as a black line in the upper panel.
The central panel shows an example of dataset (black line) made of a Gaussian white noise 
with $\sigma=30$ns plus the injected signal (red line). The lower panel show the residual noise
after the best estimate of the signal was removed. The residual signal after removal of the 
best estimate is shown as a red--dashed line in the upper panel.}
\label{F:timedomain}
\end{figure}

We finally tackled a more `realistic' situation in which white noise was added to the data (note that the 
spectral properties of the noise is irrelevant here, since we perform a `monochromatic' exercise).
We considered two situation representative of realistic scenarios that might be realized in the near future:
\begin{enumerate}
\item {\it Hard search}: 30 pulsars with equal rms noise of 100ns, which is the nominal sensitivity goal 
of the IPTA collaboration \cite{hobbs2010};
\item {\it Easy search}: 50 pulsars with equal rms noise of 30ns, which is what it might (conservatively) be
achieved with the SKA.
\end{enumerate}
Both datasets were again equally sampled once every two weeks over a timespan of 10 years, and few sources 
with unequal strength of the order of $50-100$ns (consistent with the strongest sources predicted by 
state of the art SMBHB population models) were injected at a frequency $f=2\times10^{-8}$Hz. 
An example of `easy' data stream is shown in figure \ref{F:timedomain}. The signal is
clearly visible by eye in the noisy data (middle panel). In the `easy' search, SNRs in each single pulsars
were in fact between 5 and 30, with total SNR in the array of $162.33$. In the `hard' case, SNRs in 
some individual pulsars were even $<1$, with a total SNR in the array of $23.65$.

\begin{figure*}
\centering
\begin{tabular}{cc}
\includegraphics[width=0.35\textheight, keepaspectratio=true]{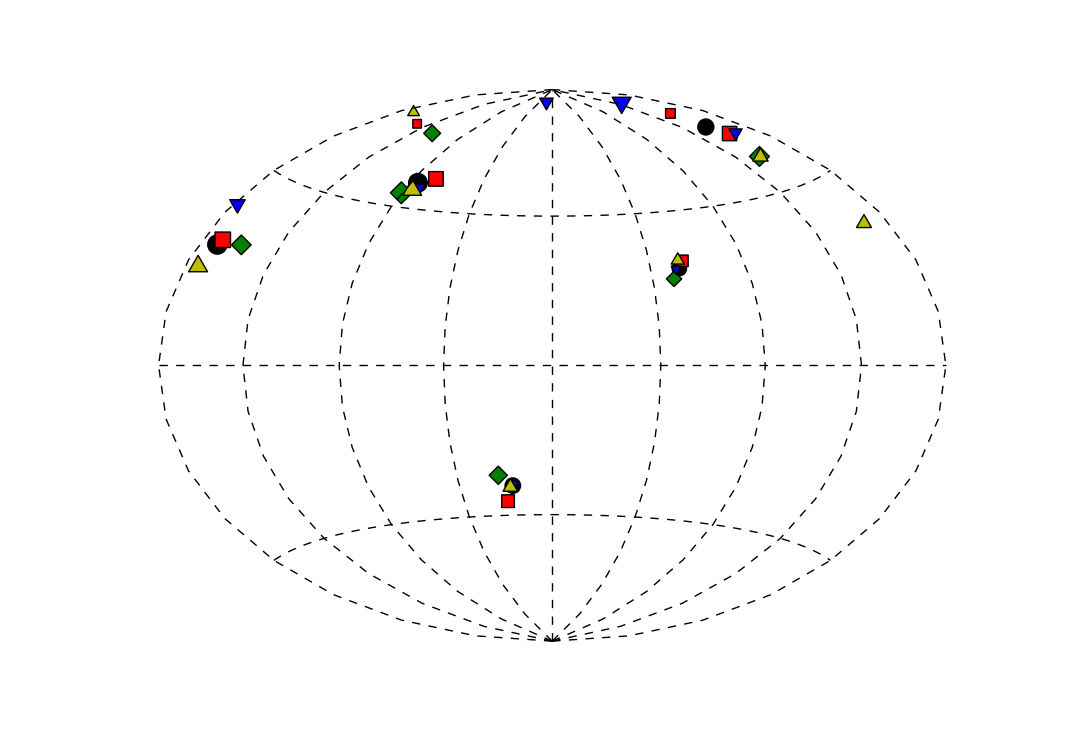} &
\includegraphics[width=0.35\textheight, keepaspectratio=true]{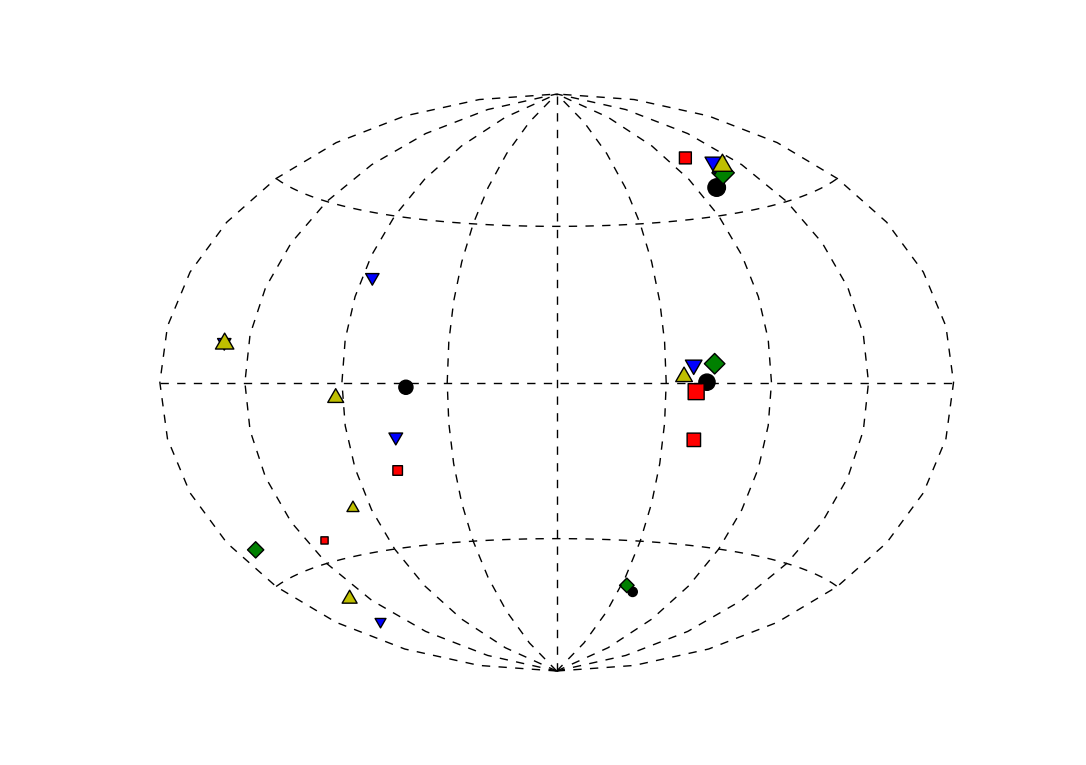}\\ 
\end{tabular}
\caption{Results of the blind search with noise. Left panel: 'easy search'; filled--black circles 
represent the injected sources, whereas each set of colored symbols represents a 'high likelihood' solution
of the blind search. Right panel: same as the left panel but for the 'hard search'. The size of the
black dots is proportional to the injected source strength; the size of the colored symbols
is proportional to their contribution to the likelihood.}
\label{F:Blindnoise}
\end{figure*}

Results for the `easy' search are shown in the left panel of figure \ref{F:Blindnoise}, where the four
best solutions are plotted. All the solutions inferred 5-to-7 sources, pinning down the position of 
most of the injections within just a few degrees. In particular, the seven-source solution showed
with red squares recovered correctly all the five sources within $<10$ degrees of their true location, 
whereas the two additional sources inferred by the solution contribute much less (factor of two-to-three) 
to the likelihood of the solution with 
respect to the other five. A good performance was expected in this case,
since the SNR is so high that we are approaching the noiseless limit.

The four best solutions to the `hard' search are shown in the right panel of figure \ref{F:Blindnoise}.
Here we see a different situation: all the solutions correctly recover the two brightest sources within
few degrees of their true location. However the weakest two are often missing or misidentified. Four-five
source solutions are clearly preferred (higher maximum likelihood), but the weakest sources are not
correctly recovered. This is simply because their  SNR in the array ( $\sim 4.5$ and $\sim 10.1$) is both 
small in absolute terms, and relative to the brighter 
sources (SNR $\sim 13.7$ and $\sim 15.9$). 
The noise contribution is rather significant in this case, as it creates additional (to the noiseless, intrinsic)
  ambiguity by promoting secondary likelihood maxima.
 By inspecting individual datasets as the one shown in 
figure \ref{F:timedomain}, we noticed that no more than 10-15 pulsars significantly contribute to the
total SNR in the array, being the SNR in the others of order unity. This means that we effectively
have barely enough pulsars to estimate all the parameters of the four sources, which might also contribute to
the difficulties of locating the weaker sources in the sky. We expect a proper search to 
improve these results.  

\section{Discussion}
\label{discussion}

We have investigated for the first time the possibility of exploiting the information enclosed in the 
spatial distribution of a pulsar timing array to resolve gravitational wave sources individually. 
To this purpose, we employ  a maximum likelihood-based algorithm and tested its performance in a 
number of simple, idealized situations. Under the simplifications and assumptions listed in the 
Section~\ref{S:objectives}, we can draw the following conclusions:

\begin{itemize}
\item to estimate the sky location of \emph{all} $N$ GW source we need at least $P=3\times N$ 
pulsars;
\item if $P<3\times N$ we can still make reliable estimations of 
the position of some (usually the strongest) GW sources;
\item in this latter case, removing the stronger sources from the solution, also removes
power belonging to the weaker ones, causing their misidentification or misplacement;
\item it is beneficial to run low $N$-source models even if we know that there are many more 
sources. Our results show that such low $N$-source models usually allow detection and correct sky location
of the brightest GW sources; 
\item computation of the likelihood with increasingly number of sources assumed in the model 
resolves more GW sources but at the same time increases the ambiguity (error) regions;
\item if the GW sources are located close to each other in the sky, they cannot be reliably resolved 
individually, and appear as a single large ambiguity region. However, the maximum likelihood estimator
usually favors multi source models, giving valuable information on the number of sources present in 
the cluster;
\item in the limit of low number $N$ of GW sources, the maximum likelihood estimator allows to infer 
the correct number of sources present in the datasets (as long as $P>3\times N$). 
In the noiseless case, even by using very primitive search techniques, we are able
to correctly identify and locate all the sources in the sky within few degrees of their true location;
\item in presence of noise, we can also infer the correct number of sources and locate the bright ones 
in the sky. The presence of dimmer sources can be correctly inferred, but due to their low SNR they
are easily misplaced.
\end{itemize}

This study mostly represents a proof of concept of the possibility of utilizing the spatial information
enclosed in the array to disentangle signals that would otherwise appear like a confusion foreground. 
In particular, most of the results were derived in the noiseless limit. On the other hand, the employed
search methods were very simple and largely non-optimal, which somehow limited the capabilities of the
analysis. In future investigations, we plan to expand our analysis in several ways to 
assess its full potential in more realistic scenarios. In particular we will:

\begin{enumerate}
\item test the limit of large number of sources, where $N \gg P$; 
\item draw the source strength distribution from state of the art models of the cosmological
population of SMBHBs;
\item include the pulsar terms in the signal and perform frequency searches;
\item consider different noise levels and noise spectra in each single pulsar in the array;
\item optimize the effectiveness of our analysis technique by employing proper multimodal search 
algorithms, similar to those applied in \cite{Petiteau:2010zu, Feroz:2009de}.
\end{enumerate}

Pulsar timing arrays will be powerful gravitational wave detectors, and further studies
along the line of the present investigation will help exploiting their full potential as 
astrophysical observatories.
\section*{Acknowledgments}

We thank K.J. Lee for useful discussions which led to this work.  S.B. also would like 
to thank Norbert Wex for suggestions on the presentation of the results. Finally, 
authors would like to thank A. Petiteau and J. Gair for the very useful discussions.

\bibliographystyle{h-physrev4}
\bibliography{references}

\end{document}